\documentclass[aps,pre,floats,floatfix,twocolumn,longbibliography,superscriptaddress]{revtex4-1}

\usepackage{color} 
\usepackage{latexsym}
\usepackage{amsmath} 
\usepackage{amssymb} 
\usepackage{bm}
\usepackage{times}
\usepackage{float}

\usepackage{graphicx}
\usepackage{epstopdf}
\graphicspath{{figs/}}

\usepackage[colorlinks,linkcolor=blue,urlcolor=blue,citecolor=blue,bookmarks=true,pdftitle={}]{hyperref}

\newcommand{\bra}[1]{\left\langle #1 \right|}
\newcommand{\ket}[1]{\left| #1 \right\rangle}

\newcommand{\avg}[1]{\left\langle #1 \right\rangle }

\newcommand{\beq}{\begin{eqnarray}}
\newcommand{\eeq}{\end{eqnarray}} 

\newcommand{\hide}[1]{}

\begin{document}

\title{Classical route to ergodicity and scarring in collective quantum systems}
 
\author{Sudip Sinha}
\affiliation{Indian Institute of Science Education and Research Kolkata, Mohanpur, Nadia 741246, India}

\author{Sayak Ray}
\affiliation{Physikalisches Institut, Universit\"at Bonn, Nu\ss allee 12, 53115 Bonn, Germany}

\author{Subhasis Sinha}
\affiliation{Indian Institute of Science Education and Research Kolkata, Mohanpur, Nadia 741246, India}
 
\date{\today}

\begin{abstract}
Ergodicity, a fundamental concept in statistical mechanics, is not yet a fully understood phenomena for closed quantum systems, particularly its connection with the underlying chaos. In this review, we consider a few examples of collective quantum systems to unveil the intricate relationship of ergodicity as well as its deviation due to quantum scarring phenomena with their classical counterpart. A comprehensive overview of classical and quantum chaos is provided, along with the tools essential for their detection. Furthermore, we survey recent theoretical and experimental advancements in the domain of ergodicity and its violations. This review aims to illuminate the classical perspective of quantum scarring phenomena in interacting quantum systems.
\end{abstract}

\maketitle

\section{Introduction}
Ergodicity is one of the central concepts of statistical mechanics \cite{Sinai-book,Cornfield-book,Halmos-book}, however, a detailed understanding of its origin is not yet clear, particularly in quantum systems. It is a general belief that chaos in a classical system with many degrees of freedom leads to ergodic time evolution. According to the ergodic hypothesis, the system can thereby explore the entire available phase space, which results in an equivalence between the time averaging and ensemble averaging of the macroscopic thermodynamic quantities \cite{Boltzmann}. This also leads to the emergence of a thermal steady state describing the equilibrium, where the memory of the initial conditions is lost. 
While the ergodic hypothesis generally applies to non-integrable systems with non-linearity, it's important to note that non-linear interactions alone may not always lead to ergodic behavior. The Fermi-Pasta-Ulam-Tsingou (FPUT) problem is a notable example of this departure from ergodicity \cite{FPU, Izrailev-rev-FPU, Tsingou}. 

On the contrary, ergodicity in a closed quantum system is a priori not explicit, since the concept of phase space chaos is absent due to the Heisenberg uncertainty principle. 
The first crucial step towards understanding the problem of quantum ergodicity was taken by von Neumann, who investigated the `quantum ergodic theorem', which focuses on the distribution of the macroscopic observables and its correspondence with the microcanonical ensemble in the long run \cite{Neumann_quantum_ergodic_theorem1929,Goldstein2010}.
In recent years, this issue has regained interest in the context of {\it thermalization} of isolated quantum systems \cite{Polkovnikov-review}, for which the system relaxes to a steady state during the time evolution, starting from an arbitrary state.
The expectation value of the local observables in such a dynamically evolved steady state attains an equilibrium value, which can be described by an appropriate ensemble of statistical mechanics. In the presence of additional conserved quantities, the thermal steady state corresponds to the `generalized Gibbs ensemble' (GGE) \cite{Rigol-GGE,Rigol-GGE2,Caux-GGE}. 
During the non-equilibrium dynamics, the emergence of such equilibrium in the asymptotic regime is commonly known as `thermalization' and has been observed experimentally in cold atom setups \cite{Ueda-rev-2020,Relaxation_correlated_bosons2012,GGE_experiment_Langen2015,
Thermalization_through_entanglement2016,
Relaxation_Rydberg_spinsystem2018,Lev2018,
Relaxation1_lattice_gauge_simulator2023}.
To understand thermalization in a quantum system from a microscopic perspective, the eigenstate thermalization hypothesis (ETH) has been proposed, according to which thermalization occurs at the level of individual eigenstates \cite{Deutsch1991, Srednicki1994, Srednicki1999, Rigol2008, Polkovnikov-review, Santos-review, ETH_review_Deutsch2018}.
More precisely, ETH suggests that the statistical properties of the eigenstates can play an important role in thermalization.
Typical eigenstates satisfying ETH also follow Berry's conjecture which states that higher-energy eigenstates can be decomposed into random components (either Fourier components or site amplitudes) with Normal distribution \cite{Berry-conjecture, Jarzynski1997}. Statistical properties of such eigenstates are similar to eigenvectors of a random matrix.
This remarkable finding opens up broader avenues for investigating quantum ergodicity through the utilization of Random Matrix Theory (RMT). Importantly, RMT is not tied to specific systems, offering a more generalized approach to the study of quantum ergodicity.

Historically, RMT was introduced to capture the statistical properties of the energy spectrum of complex quantum systems, particularly of atomic nuclei \cite{Wigner_original1,Wigner_original2,Wigner_original3,Mehta-RMT, Brody-RMP}. Later on, the Bohigas-Giannoni-Schmit (BGS) conjecture \cite{BGS} led to an important development, which built the basic pillar for quantum signature of chaos. According to this conjecture, the spectral properties of a Hamiltonian, which undergoes a classically chaotic dynamics, follow the predictions of RMT ensembles \cite{Haake2010,Stockmann_book}. Such a connection has been explored in a variety of systems including those which undergo a transition from regular to chaotic dynamics \cite{Haake2010,Stockmann_book}. 
Naively, BGS conjecture builds a bridge between classical chaos and the random eigenstates of its quantum counterpart required for quantum ergodicity and ETH, where RMT plays a central role in characterizing the degree of randomness. Thus, the RMT has turned out to be a powerful tool not only to diagnose chaos but also to investigate the ergodic behavior and thermalization process of a generic many-body system \cite{Polkovnikov-review,Santos-review,Bohigas1975,Pandey1979,
Bohigas_chaos_and_RMT,Ullmo1993, Zelevinsky1996, Weidenmuller1998, Weidenmuller2000,Lebouef2000, Ueda2021}. 
In this review, we will elucidate such a connection between the ergodicity of the isolated quantum systems and chaos using collective systems with appropriate semiclassical limit, which allows us to study the corresponding dynamical behavior directly from the underlying phase space of the system. 

In recent years, another technique, namely the out-of-time order correlator (OTOC), has been introduced to diagnose chaos in a generic interacting quantum system, which is also referred to as `many-body quantum chaos' \cite{Shenker2014, Maldacena2016, Swingle-QFT-OTOC, Galitski2017, Sachdev2017, Knap2017,Swingle_unscrambling2018, information_scrambling2020,information_scrambling_finite_temp,OTOC_anisotropicDicke, quantum_ising_chain2018,finite_speed_scrambling,OTOC_classicalquantumDicke, quantum_classical_crossover,OTOC_KDM,Swingle2016,Lev2017,Galitski-levstat,GarciaMata-OTOC,GarciaMata-OTOC-fluc,Richter_saturation,markovic_saturation,Arul-scrambling,Fradkin-OTOC,Ray-OTOC-MBL,Alet-OTOC,RMT_isospectral_twirling,Clifford_gates_quantum_chaos,p-spin_glass}.
The growth rate of OTOC, which was first introduced in the context of disordered superconductor \cite{Larkin}, can play an analogous role to the Lyapunov exponent in determining the degree of chaos \cite{Shenker2014, Maldacena2016, Swingle-QFT-OTOC, Galitski2017,Sachdev2017,Swingle_unscrambling2018, information_scrambling2020, GarciaMata-OTOC, OTOC_classicalquantumDicke, quantum_classical_crossover, OTOC_KDM, Galitski-levstat,Clifford_gates_quantum_chaos,p-spin_glass}, that has been popularized after its application to black hole thermalization \cite{Shenker2014, Maldacena2016}, as well as its connection to quantum information scrambling \cite{Swingle2016, Lev2017, finite_speed_scrambling, Knap2017, quantum_ising_chain2018, information_scrambling2020, information_scrambling_finite_temp, RMT_isospectral_twirling}. Although, this correspondence is not very obvious in generic interacting quantum systems, nevertheless the dynamics of OTOC can provide useful information, particularly regarding the ergodicity of the system \cite{GarciaMata-OTOC,OTOC_KDM,OTOC_anisotropicDicke, Galitski-levstat, Richter_saturation, Arul-scrambling, GarciaMata-OTOC-fluc, markovic_saturation, Fradkin-OTOC, Ray-OTOC-MBL, Alet-OTOC}. 
The scrambling of information in a complex quantum system is an interesting phenomena by itself, and its connection with ergodicity can also be diagnosed by measuring the growth of entanglement.

Apart from the onset of chaos and ergodicity in a closed quantum system, understanding its deviation has also become an intense area of research, which is attributed to the ergodicity breaking phenomena.
Contrary to the general expectation, not all interacting quantum systems exhibit ergodic time evolution despite their complexities.
The commonly known example is the dynamics of a system in the presence of disorder, the concept of which was introduced by P. W. Anderson, where non-interacting particles undergo localization \cite{Anderson-loc}. While, an interacting system can exhibit a many-body localized (MBL) phase \cite{Basko-MBL, MBL_quasiperiodic2013, Huse-rev-MBL, Abanin2019}, which strongly violates ergodicity and has been realized experimentally in the presence of correlated disorder, namely the quasiperiodic potential \cite{Roati-qp, Schreiber2015, Choi2016, DeMarco2009}.
In addition, `glassy' phases such as spin glasses can also appear in frustrated and disordered systems, which break ergodicity \cite{spin_glass_review1, spin_glass_review2,Langer_spin_glass_review, ergodicity_breaking_metric_classical,quantum_glass_book,
quantum_glass_review}.
Ergodicity breaking can also occur in the absence of disorder. For example, in the cold atom experiments it is revealed that certain many-body systems, such as one-dimensional (1D) arrays of coupled Bose-Einstein condensates (BEC), forming a quantum Newton's cradle, fail to attain equilibrium in the long time \cite{Kinoshita2006}. 
In a recent experiment, it has been observed that some specific initial states in an array of strongly interacting ultracold Rydberg atoms do not thermalize and exhibit long-time coherent oscillations, which is in contrast to the other initial states that undergo an ergodic time evolution in the same parameter regime \cite{Rydberg_expt_scar}. Such athermal behavior due to the presence of certain atypical states in the ergodic regime leads to a weak breaking of ergodicity, which have been dubbed as `many-body quantum scars' (MBQS) \cite{Abanin_Rydberg_scar2018,Papic_Rydberg_scar2018, Abanin_scars_review2021, Moessner_scars_review2022, Moudgalya_review2022}.
The retention of memory of the initial states leading to revival dynamics due to a vanishingly small fraction of athermal states in the ergodic regime is a defining feature of weak ergodicity breaking \cite{Abanin_scars_review2021}. 
Motivated by this experiment, a huge impetus has been generated to theoretically study the mechanism behind the formation of MBQS in different models, particularly in the `PXP' chain \cite{Abanin_Rydberg_scar2018, Papic_Rydberg_scar2018, Motrunich_Exact_Scar_Rydberg2019, Khemani_Rydberg_integrability2019, scar_Papic2020, magnonscars2020, scars_Motrunich2020, scar_correspondence_principle2021,
Choi_Emmergent_Symmetry2019, Lukin_Periodic_Orbits2019, Abanin_PRX2020}, Affleck-Kennedy-Lieb-Tasaki (AKLT) model \cite{AKLT_scar_Moudgalya1,AKLT_scar_Moudgalya2,AKLT_scar_Motrunich2020,AKLT_scar_MPS2020,AKLT_Shiraishi2019}, spin-1 XY magnets \cite{spin1_XY_scars2019,spin1_XY_scarsMPS2020}, Hubbard models \cite{eta_pairing_Moudgalya2020,eta_pairing_Motrunich2020, correlated_hopping_BHM2020, scars_optical_lattice2020, tilted_1d_FHM2021, tilted_1d_FHM2023, Richter2023, bilayer_scars2022}, and many more \cite{scar_Mori2017, Onsager_scars2020, Fracton_scars2020, kagome_scars2020, transverse_ising_ladder2020, lattice_gauge_scars2021, lattice_gauge_scars2022a, lattice_gauge_scars2022b, scars_deformedalgebra2020, rainbow_scars2022, multiple_magnons2022, Fradkin_chains2022, Kitaev_chain2022, Kitaev_chain2023, truncated_schwinger2023a, truncated_schwinger2023b, Heisenberg_clusters2023,Evrard}. 
Apart from the Rydberg quantum simulator, the MBQS have also been observed experimentally in other platforms like superconducting qubits \cite{scars_scqubits2022} and Bose-Hubbard quantum simulator \cite{BHM_simulatorscar2022}. 
It is worth mentioning that, in terms of weak breaking of ergodicity,  MBQS bears resemblance with the `quantum scars' of single particle chaotic systems \cite{Heller1984,Kaplan1998,KaplanPRE1999,Kaplan1999}. Originally, the `quantum scars' have been identified as a reminiscence of the unstable orbits in the wavefunctions of a chaotic system, namely the billiard \cite{Heller1984}. From this point of view, the underlying classical picture of MBQS is not fully understood due to the absence of a suitable classical limit of a generic many-body system, which deserves further investigation. In this review, we will shed some light on the issue of the formation of scars and its connection with the underlying classical dynamics of collective models. Other than the MBQS, there can be other sources of ergodicity violation, such as the presence of non-ergodic extended states exhibiting multifractality \cite{Anderson_transitions_Mirlin2008, multifractal_spinchains_Bogomolny2012, Altshuler2014, Altshuler2016, Luitz2016, drive_induced_AA2016, Luitz2015, ShankarDasSarma2015, Santos2015,  Abanin2017, Santos_nonergodic2017, Lindinger2019,Nicolas2019, powerlaw_hopping_deng2019, Khaymovich2019, Mondaini2019, Khaymovich2020, Luitz_transport2020, Fava2020, Subroto2020, Buchleitner2021, Sumilan2023}. 
A disorder-free system can also show slow thermalization and non-ergodicity due to kinetic constraints \cite{Cirac2020, Sthitadhi2022}. Furthermore, in the presence of constraints, an interacting quantum system can exhibit `Hilbert space fragmentation', which can even lead to a stronger violation of ergodicity \cite{Pollmann2020,Nandkishore_fragmentation2020,Krylov_fracture_book,
Motrunich_PRX2022,non_ergodicity_tilted_FHM2021,Moudgalya_review2022}.

\begin{figure*}
	\centering
	\includegraphics[width=1.02\textwidth]{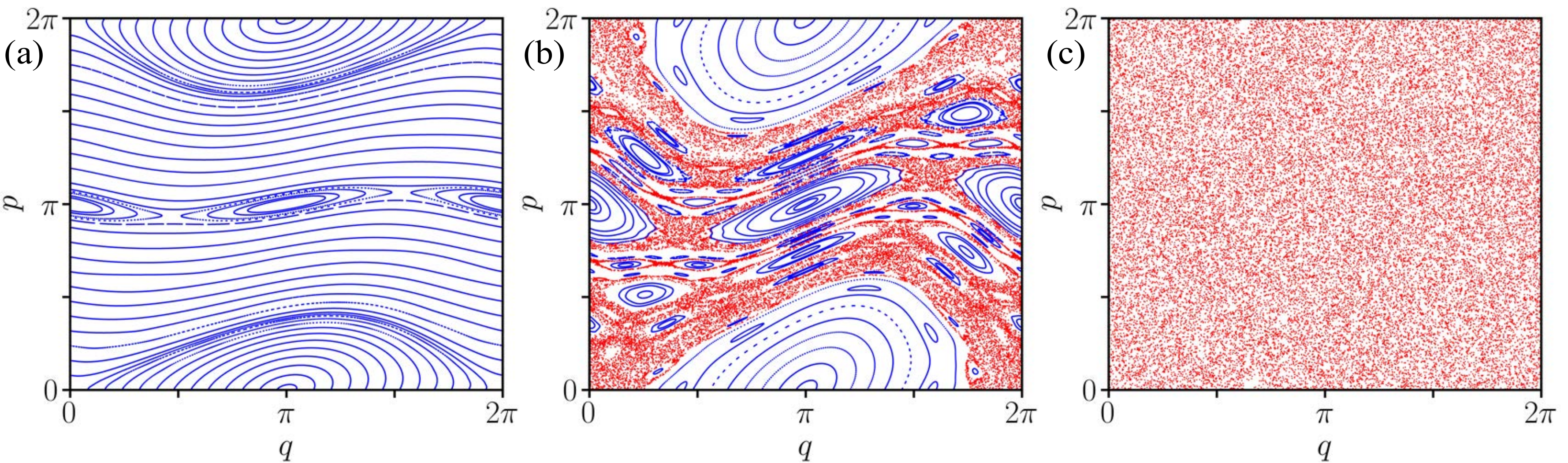}
	\caption{Evolution of Phase space behavior of the classical kicked rotor with increasing kicking strength $\kappa$: (a) Regular behavior at $\kappa = 0.4$, (b) mixed phase space behavior with the coexistence of regular (blue) as well as chaotic (red) trajectories at $\kappa = 1.0$, and (d) fully chaotic dynamics at $\kappa = 7.0$. For more details regarding the kicked rotor, see Ref.\cite{Review_KR}.}      
\label{Fig1}
\end{figure*}

This review serves dual purposes as we state below.\\ 
(I) We elucidate the concept of ergodicity in isolated quantum systems and its breaking caused by phenomena like scarring. We explore their relationship with the underlying classical behavior using specific examples from the collective quantum systems, which are realizable in cold-atom experiments.\\ 
(II) A concise overview of quantum chaos, ergodicity, as well as its deviation, and the recent developments in these areas are presented.\\ 
The review is organized as follows. In Sec.~\ref{dynamical_systems_and_chaos}, we provide a pedagogical description of the classical dynamics and chaos in the Hamiltonian systems. In the next two sections, namely Sec.~\ref{eigenvalue_statistics} and Sec.~\ref{eigenvector_statistics}, we briefly discuss the methods and tools to diagnose the quantum signature of chaos. The concept of thermalization is presented in Sec.~\ref{ergodicity_and_thermalization}. In Sec.~\ref{quantum_classical_correspondence_and_collective_models}, we introduce the different collective models to explore the classical-quantum correspondence and their connection with ergodicity. In the following sections, we discuss the many-body quantum chaos from out-of-time-order correlator (in Sec.~\ref{OTOC}) and from entanglement (in Sec.~\ref{chaos_and_entanglement}). The deviation from ergodicity due to scarring phenomena and the energy-dependent ergodic behavior is described in Sec.~\ref{deviation_from_ergodicity}. The signature of the underlying classicality in terms of the entanglement spectrum is discussed in Sec.~\ref{classicality_and_entanglement_spectrum}. Finally, we present the discussion and outlook in Sec.~\ref{discussion_outlook}.

\section{Dynamical systems and chaos}
\label{dynamical_systems_and_chaos}
The paradigm of chaos in dynamical systems has revolutionized our understanding in different areas of science, particularly in statistical mechanics \cite{Ruelle_stat_mech_book, Ruelle1985, Zaslavsky_chaos_book}, which has regained interest in recent years in the context of ergodicity of a closed quantum system \cite{Polkovnikov-review, Santos-review}.
In general, the behavior of dynamical systems can be categorized into the following two broad classes: regular and chaotic. Usually, the integrable systems exhibit regular motion due to the presence of conserved quantities, which are known as integrals of motion. Whereas chaotic dynamics is observed in non-integrable systems, for which the trajectories exhibit exponential sensitivity towards the initial conditions.

A dynamical system comprised of $N$ degrees of freedom can be described by a set of generalized coordinates $\{q_{i}\}$ and the corresponding conjugate momentum $\{p_{i}\}$ with $i=1,..,N$, which define its phase space. In this review, we only consider the dynamical systems which satisfy Hamilton's equation of motion,
\begin{eqnarray}
\dot{q}_{i} = \frac{\partial \mathcal{H}}{\partial p_{i}};\quad \dot{p}_{i} = -\frac{\partial \mathcal{H}}{\partial q_{i}}
\end{eqnarray}
where $\mathcal{H}$ is the Hamiltonian of the system. For integrable  systems, there exist $N$ integrals of motion $\{I_i\}$, satisfying the Poisson bracket relation $\{\mathcal{H},I_{i}\}_{\rm PB}=0$, due to which the motion is restricted on $N$ dimensional tori \cite{Zaslavsky_chaos_book, Wimberger_book, Arnold_book}. Such a large number of conserved quantities are the key ingredients for regular phase space dynamics. On the contrary, chaotic dynamics is observed in non-integrable systems, where two initial phase space points with infinitesimally small separation $\delta \boldsymbol{X}$ diverge exponentially with time, $||\delta \boldsymbol{X}(t)|| \sim e^{\lambda_{\rm L} t} ||\delta \boldsymbol{X}(0)||$, where $\boldsymbol{X}=\{q_{i},p_{i}\}$ and $||.||$ is the Euclidean norm. Such exponential sensitivity towards the initial conditions is the main characteristic feature of chaotic dynamics and is known as the  `butterfly effect'. The degree of chaos can be measured from the Lyapunov exponent (LE) $\lambda_{\rm L}$ defined as \cite{Strogatz},
\begin{eqnarray}
\lambda_{\rm L} = \lim_{t \to \infty} \lim_{||\delta \boldsymbol{X}(0)|| \to 0} \frac{1}{t} {\rm ln}\left(\frac{||\delta \boldsymbol{X}(t)||}{||\delta \boldsymbol{X}(0)||}\right),
\end{eqnarray}
This is the most secular method to quantify the degree of chaos in dynamical systems and its practical computation can be done using different numerical techniques \cite{Lauterborn1990}. However, in recent years, different theoretical frameworks have been introduced to serve this purpose, particularly in many-particle systems, which we will discuss in the later part of this review.

In the absence of the constraints imposed by the conserved quantities, the dynamical system can explore the available volume of phase space, leading to the `phase space mixing'. Such dynamical behavior is believed to be responsible for ergodicity, which serves as one of the basic pillars of statistical mechanics.
Ergodicity for a classical system is a well defined concept, according to which, the system traverses over all the possible configurations of the phase space when given enough time \cite{Ruelle1985}. Consequently, the time average of a thermodynamic quantity becomes equivalent to its ensemble average over an appropriate phase space distribution.
Although the connection between chaos and ergodicity is not straightforward and well established, it is however a mathematically involved topic, for which we refer the interested readers to Refs.\cite{Sinai-book, Cornfield-book, Halmos-book}. We would like to point out that, there are certain cases in which the system explores the phase space over time, however without the exponential sensitivity on initial perturbation i.e. $\lambda_{\rm L}=0$. This class of dynamical systems neither belongs to regular nor to chaotic systems, they are known as strange non-chaotic attractors, which exhibit fractal structures and can be found in quasi-periodically driven systems with incommensurate frequencies \cite{Pikovsky1994, Orland1988, Pikovsky1996, Lebowitz1992, Geisel1990, Ramaswamy2010, Ott1990}. Such a dynamical behavior can have a great impact on the ergodicity of a system in the presence of interaction \cite{Potter2018, Ray-quasi-Floquet, Nandy2017, Nandy2018, Sen2019}, which we will also see in different contexts as we proceed further in this review.

Interestingly, there is a class of dynamical systems, that exhibit a transition from regular to chaotic dynamics, by tuning some parameters of the Hamiltonian. Particularly, one can identify an intermediate regime, where regular and chaotic trajectories can coexist, resulting in mixed phase space dynamics. This type of system has attracted a lot of attention in the past few decades and has revealed different kinds of interesting dynamical behavior. Although the pendulum is the simplest example of an integrable system with regular phase space trajectories, it can undergo a transition to chaos under a periodic time-dependent perturbation. Such a transition to chaos in a periodically driven system has been extensively studied in paradigmatic models like the kicked rotor \cite{Chirikov_KR,Casati_KR,Review_KR} as well as kicked top \cite{Haake1987,Haake2010}. Here we point out that, although the phenomena of chaos is absent in one-dimensional time-independent systems, the time-dependent perturbations can give rise to chaotic behavior, which can occur by tuning the strength of the periodic perturbation as evident from the change in the phase space behavior of the kicked rotor model shown in Fig.\ref{Fig1}.

The onset of chaos, particularly its transition from an integrable regime, is a mathematically involved topic. In this context, the Kolmogorov–Arnold–Moser (KAM) theorem sheds light on the deformation and breaking of the invariant tori, as the perturbation strength increases \cite{Arnold_book}. 
The mixed phase space in the intermediate regime of perturbation strength, consisting of regular islands within the chaotic sea of higher entropy, is fascinating and deserves further attention. The local variation of chaoticity which, although can be visualized from the phase space, however, is quantified from the Kolmogorov-Sinai entropy (KSE) \cite{KSE1, KSE2}, that is also related to the Lyapunov exponent \cite{Pesin1977}. The mixed phase space dynamics raises interesting questions, such as its manifestation in the corresponding quantum system and its role in the overall ergodicity of the system.   
In the presence of interactions, such correspondence can be elucidated in the collective models with well-defined semiclassical limit, which we will elaborate in the subsequent sections of this review.

It is worth noting that, there are several instances when the system is restricted from exploring the entire phase space, which can lead to the breaking of ergodicity. In particular, this occurs in the context of symmetry breaking, which causes a phase transition, and as a result, the configuration space is fragmented into smaller regions \cite{Goldenfeld_book,Palmer_ergodicity_breaking}. Furthermore, some systems exhibit `glassy' behavior at low temperatures due to the presence of a huge number of local equilibria in the free energy landscape generated by frustration or disorder \cite{spin_glass_review1,spin_glass_review2,Langer_spin_glass_review,
ergodicity_breaking_metric_classical}. Consequently, the system gets stuck in one of these configurations, leading to relaxation times significantly longer than experimentally observable time scales, thereby resulting in ergodicity breaking. In addition, such phenomena has also been observed in recent experiments involving artificial spin ice \cite{spin_ice_expt1,spin_ice_expt2}.
 
\section{Quantum signature of chaos from Random Matrix Theory}
\label{section-RMT}
In a quantum system, the notion of a phase space trajectory is absent, and therefore, unlike in a classical system, the exponential sensitivity to a small change in the initial condition, which is the hallmark of chaos, is not directly applied. Thus, finding the quantum signature of chaos attracted much attention in the past years. A possible connection came through the correlated energy spectrum of a classically chaotic system, where the energy levels exhibit level repulsion \cite{BGS}. The degree of energy level repulsion in such correlated spectra depends on the underlying symmetries of the quantum system, and can be understood using the Random Matrix Theory (RMT) \cite{Mehta-RMT,Haake2010,Stockmann_book}. This connection with the eigenvalue and eigenvector statistics is briefly discussed in the next subsections. In recent years, such a connection has regained interest, particularly in the context of quantum ergodicity and thermalization, which we will demonstrate with few examples from the collective models.

\begin{figure}
	\centering
	\includegraphics[width=\columnwidth]{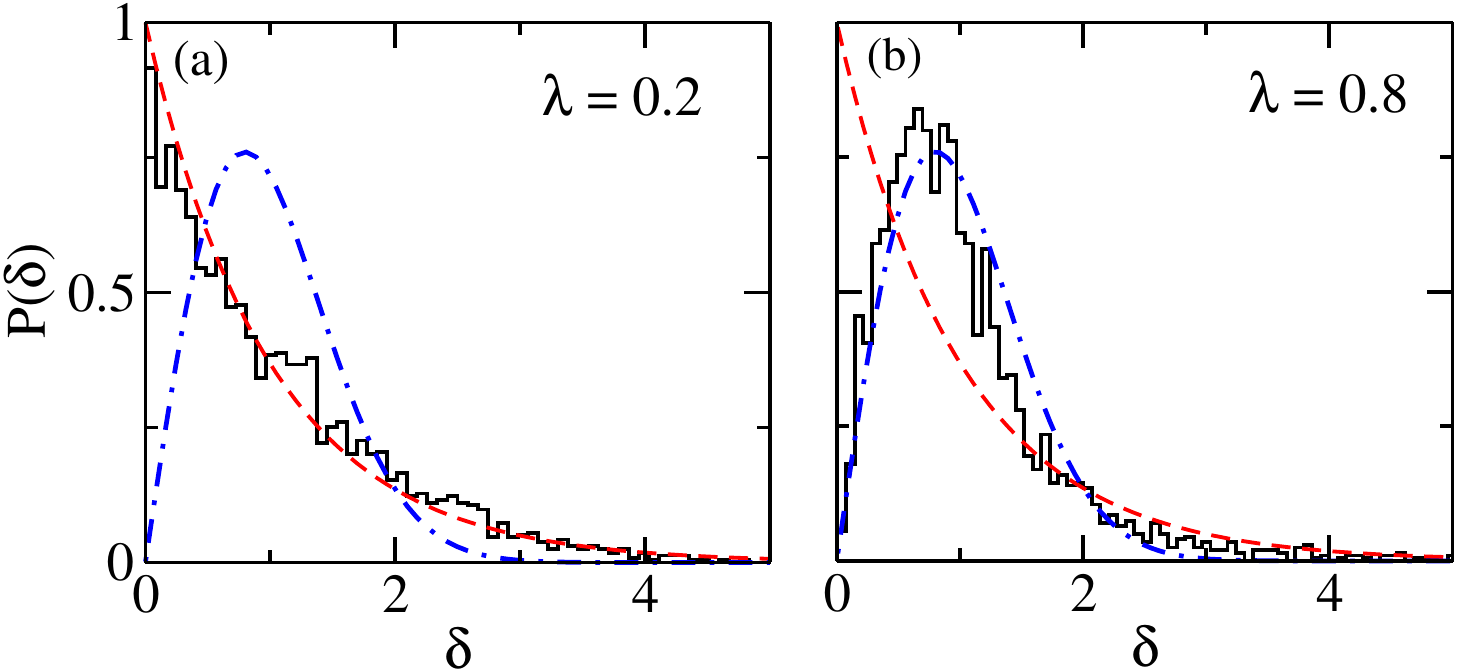}
	\caption{Plots of nearest-neighbor level spacing distributions $P(\delta)$ of the Dicke Hamiltonian in Eq.~\eqref{Dicke_model} for different coupling strengths $\lambda$ and spin $S=25$. The dotted and dashed lines denote the universal Poissonian and Wigner 
 (GOE) distributions, respectively. The Hamiltonian is on scaled resonance: $\omega=\omega_0=1$, $\lambda_c=0.5$.}      
\label{Fig2}
\end{figure}

\subsection{Eigenvalue statistics}
\label{eigenvalue_statistics}
A non-integrable quantum system usually displays correlated spectra, where the energy levels exhibit a certain degree of level repulsion. In the seminal work on characterizing the universality of such energy level statistics of a classically chaotic system, Bohigas, Giannoni, and Schmit proposed the `BGS conjecture', which is till date used as a widely applicable link between chaos and its quantum signature. More precisely, according to the BGS conjecture, the statistical properties of the energy levels in a quantum system, whose classical counterpart exhibits chaotic dynamics, follow the predictions of RMT \cite{BGS}. Depending on the symmetry of the Hamiltonian (e.g. time-reversal symmetry), the energy level spacing distribution of the chaotic spectra can be broadly classified into three Gaussian ensembles of the RMT given by \cite{Mehta-RMT,Haake2010},
\begin{equation} 
P(\delta)= \begin{cases} 
      (\delta \pi/2)e^{-\delta^2 \pi/4} & \text{orthogonal} \\
      (\delta^2 32/\pi^2)e^{-\delta^2 4/\pi} & \text{unitary} \\
      (\delta^4 2^{18}/3^6 \pi^3)e^{-\delta^2 64/9 \pi} & \text{symplectic}
      \end{cases}
\end{equation}
The three ensembles differ by the degree of level repulsion, $P(\delta) \sim \delta^\beta$, where the Dyson index $\beta = 1$, $2$ and $4$ correspond to GOE, GUE, and GSE of the random matrices, respectively \cite{Dyson_index}, see also Ref.\cite{Atland_classification} for non-standard symmetry classes. 
In contrast, the energy levels in the spectrum of an integrable system can cross, leading to the observation of level clustering. According to the Berry-Tabor conjecture \cite{BT_conjecture}, a Poisson level spacing distribution, $P(\delta) = e^{-\delta}$, is observed when the corresponding classical system exhibits regular phase space dynamics.
The above-mentioned spectral statistics have been widely used to distinguish the two extreme dynamical behavior, i.e. the regular motion in the integrable regime and chaotic dynamics in the presence of non-integrable perturbations. 
Such transition from regular to chaotic motion by tuning the interaction strength, and its reflection in the eigenspectrum has been nicely demonstrated in the paradigmatic Dicke model (DM) \cite{Emary_Brandes_DM, Emary_Brandes_DM-rev}, see also Fig.~\ref{Fig2} and Fig.~\ref{Fig3}. This model and its variants are discussed in Sec.~\ref{quantum_classical_correspondence_and_collective_models}, which can serve as suitable candidates to elucidate the quantum signature of chaos, due to the well-defined semiclassical limit.

More recently, an alternative quantity called the average level spacing ratio $\langle r \rangle$ has been introduced to quantify the degree of chaos \cite{D_Huse_2007_MBL,Bogomolny_2013}, which is defined as,
\begin{equation}
    \avg{r} = \avg{\rm min(\delta_n, \delta_{n+1})/max(\delta_n, \delta_{n+1})}
\label{avg_level_spacing_ratio}
\end{equation}
where, $\delta_n = \varepsilon_{n+1}-\varepsilon_n$ is the nearest neighbor level spacing for the $n$th eigenmode.  
The values of $\avg{r}$ can be computed for different spectral statistics, which are tabulated in Table~\ref{tab:ravg}. 
Note that, the values given in the table (see also Ref.\cite{Bogomolny_2013}) have been calculated analytically for random matrices of size $3 \times 3$. For larger matrices of size $N \times N$ ($N \gg 1$), there are no analytical results, however these values can be estimated numerically (see for example Ref.\cite{Luitz_note}).
\begin{table}
\begin{tabular}{|c|c|c|c|c|}
\hline
Ensemble & Poisson & GOE & GUE & GSE \\ 
\hline
$\avg{r}$ & $2\ln2-1$ & $4-2\sqrt{3}$ & $\frac{4\sqrt{3}-\pi}{2\pi}$ & $\frac{64\sqrt{3}-15\pi}{30\pi}$ \\
\hline
\end{tabular}
\caption{Values of $\avg{r}$ for different Random Matrix Theory (RMT) ensembles (see also Ref.\cite{Bogomolny_2013}).}
\label{tab:ravg}
\end{table}
This quantity serves as an analogous order parameter to quantify the degree of chaos from the eigenspectrum of the quantum system, which is useful for identifying the onset of chaos by tuning the relevant control parameters. It is worth mentioning that the spectral statistics of the localized phase follows the Poisson distribution \cite{D_Huse_2007_MBL}, which can also be employed to study the transition to delocalized phases \cite{Anderson_transitions_Mirlin2008}.

An intermediate spectral statistics, which neither belongs to Poisson nor any of the Gaussian ensembles of the RMT \cite{Brody_distribution}, can be observed in certain cases \cite{Lenz_Haake,Prosen_Robnik,Berry_Robnik,Seligman, Bogomolny_intermediate_statistics,Garcia2005,Anderson_transitions_Mirlin2008}, particularly near the metal insulator transition \cite{Anderson_transitions_Mirlin2008}.
Interestingly, systems in the mixed phase space regime can also exhibit such intermediate statistics \cite{Berry_Robnik,Seligman,Prosen_Robnik}, which can be related to the fractal nature of the spectrum, that has been observed in the kicked Harper model \cite{Geisel_metamorphosis,Casati_KHP1,Casati_KHP2,
Geisel_energy_dependence}.

Apart from the spectral statistics, the statistical properties of the eigenvectors also plays an important role in determining the ergodicity of the system, particularly for those exhibiting deviation from RMT of Gaussian class. Interestingly, such deviation can lead to non-ergodic as well as energy-dependent ergodic behavior, that is usually associated with the multifractal nature of the eigenstates, as discussed in the next subsection.
  
\subsection{Eigenvector statistics}
\label{eigenvector_statistics}
The degree of ergodicity and chaos is also embedded in the eigenvectors of a quantum system. The degree of delocalization of eigenvectors in the basis states constituting the Hilbert space can be measured by their moments defined as,
\begin{equation}
    I_q^n = \sum_{m=1}^N |\psi_n(m)|^{2q}, ~~ \bar{I}_q = \frac{1}{\mathcal{N}_n} \sum_{n\in \Delta \varepsilon_n} I_q^n \propto \mathcal{N}^{-\tau_q}
\end{equation}
where $I_q^n$ is the generalized inverse participation ratio (IPR) of the $n$th eigenstate in the basis $|m\rangle$ spanning the $\mathcal{N}$-dimensional Hilbert space and $\bar{I}_q$ is the average IPR of $\mathcal{N}_n$ number of eigenstates lying within a small energy interval $\Delta \varepsilon_n$ around eigenenergy $\varepsilon_n$. The scaling of $\bar{I}_q$ with $\mathcal{N}$ leads to obtaining the fractal dimension $D_q$ given by, $\tau_q = D_q (q-1)$ \cite{Anderson_transitions_Mirlin2008}. The fractal dimension $D_q$ can also be obtained by computing the R\'enyi entropy \cite{Renyi_original_paper} as follows,
\begin{equation}
    D_q = \lim_{\mathcal{N}\rightarrow \infty}\frac{S_q^n}{\log \mathcal{N}}, ~~ S_q^n = -\sum_{m=1}^N \frac{\log |\psi_n(m)|^{2q}}{q-1}
\end{equation}
where, $S_q^n$ denotes the R\'enyi entropy of the $n$th eigenvector. When $q\rightarrow 1$, the R\'enyi entropy $S_q^n$ becomes the same as Shannon entropy $S^n_{\rm Sh}$ for the $n$th eigenstate, and $D_{1}$ is obtained by \cite{Santos_nonergodic2017},
\begin{equation}
    D_1 = \lim_{\mathcal{N}\rightarrow \infty}\frac{S^n_{\rm Sh}}{\log \mathcal{N}}, ~~ S^n_{\rm Sh} = -\sum_{m=1}^N |\psi_n(m)|^2\log |\psi_n(m)|^{2}
\end{equation}
For the ergodic states, $D_q = 1$, while for localized states, $D_q = 0$, whereas, in the case of non-ergodic extended states, $0<D_q<1$ and $D_q$ exhibits a non-trivial dependence on $q$, which indicates multifractality of the eigenstates. 
The multifractal behavior  by itself is an interesting topic in nonlinear dynamics and we refer the readers to Ref.\cite{Mandelbrot_book} for further discussion on this.
Notably, critical states with multifractal properties can appear near the boundary separating the ergodic and non-ergodic regions of the energy spectrum \cite{Luitz2015,ShankarDasSarma2015,powerlaw_hopping_deng2019,Mondaini2019,Subroto2020,
Buchleitner2021,Sumilan2023}, giving rise to energy-dependent ergodicity. Moreover, the existence of such kind of states can directly influence the overall ergodicity of the system.

The degree of delocalization of the eigenvectors and its correspondence with the RMT can be revealed by computing the structural entropy given by \cite{structuralentropy2002},
\begin{equation}
    S_{\rm str}^n = -\sum_m |\psi_n(m)|^2 \log |\psi_n(m)|^2 + \log I_4^n, 
\end{equation}
According to the RMT prediction, for the GOE class of random matrices, the IPR can be calculated as, $I_4 = 3/\mathcal{N}$  \cite{Santos-review}, and the structural entropy attains a universal value, $S_{\rm str} \approx 0.3646$. 

Notably, the BGS conjecture hinges on a connection between the underlying dynamical chaos and the corresponding spectral properties, which in turn is related to the statistical nature of the eigenstates. This relationship can elucidate the classical route to ergodicity and thermalization in a quantum system, where RMT plays a pivotal role.

\section{Ergodicity and thermalization}
\label{ergodicity_and_thermalization}
Thermalization in an isolated/closed quantum system has remained a long-standing puzzle for the past few decades. Generally, in a non-integrable many-body system, thermalization is described by the emergence of a steady state for which the time evolution of the average of a few-body observable approaches to a steady value corresponding to an appropriate ensemble at a particular energy density. For thermalization in such closed systems, the system itself acts as its own heat bath.
Such phenomena has been tested numerically for the hardcore bosons \cite{Rigol2008}. Also, the experimental studies on the relaxation dynamics towards equilibrium in cold atom setups \cite{Ueda-rev-2020,Relaxation_correlated_bosons2012,GGE_experiment_Langen2015,
Thermalization_through_entanglement2016,Relaxation_Rydberg_spinsystem2018,Lev2018,Relaxation1_lattice_gauge_simulator2023} motivate further investigations in this direction.

Unlike the microcanonical picture of classical systems where dynamical chaos leads to ergodic behavior via phase space mixing, it is not straightforward to extend this idea in the quantum counterpart due to the absence of phase space description. 
The notion of quantum ergodicity was first explored by von Neumann in his `quantum ergodic theorem' \cite{Neumann_quantum_ergodic_theorem1929}, which was only recently touched upon in detail in Ref.\cite{Goldstein2010}. According to this theorem, for a set of typically commuting macroscopic observables, every initial wavefunction from the microcanonical energy window evolves in such a way that, the joint probability distribution of these observables obtained from the time evolved state is close to the microcanonical distribution for most of the times in the long run.
This indicates that, the statistical properties of the eigenstates play a crucial role in understanding the ergodicity of a quantum system, where RMT serves as an important tool for determining the degree of ergodicity \cite{Polkovnikov-review,Santos-review}.
The concept of quantum ergodicity is inherently related to the `thermalization' of isolated quantum systems, for which the system attains a steady state described by an appropriate thermal ensemble.
Later, to understand the route to thermalization from a microscopic point of view, the `eigenstate thermalization hypothesis' (ETH) was proposed, according to which the individual eigenstates behave like statistical ensembles in many aspects \cite{Deutsch1991, Srednicki1994, Srednicki1999, Rigol2008, Polkovnikov-review, Santos-review, ETH_review_Deutsch2018}. 
For a few-body observable $\hat{\mathcal{O}}$, its expectation value at time $t$ with respect to the time evolved state $|\psi(t) \rangle=\sum_{\alpha}C_{\alpha}e^{-\imath E_{\alpha}t} | E_{\alpha} \rangle$ ($|E_{\alpha}\rangle$ being the energy eigenstates) is given by,
\begin{align}
\langle 
\mathcal{O}(t) \rangle &= \langle \psi(t) | \hat{\mathcal{O}} | \psi(t) \rangle = \sum_{\alpha,\beta}C^{*}_{\alpha}C_{\beta}e^{\imath(E_{\alpha}-E_{\beta})t}\hat{\mathcal{O}}_{\alpha\beta} \notag\\
&= \sum_{\alpha} \hat{\mathcal{O}}_{\alpha\alpha} |C_{\alpha}|^2 + \sum_{\alpha\neq\beta}C^{*}_{\alpha}C_{\beta}e^{\imath(E_{\alpha}-E_{\beta})t} \hat{\mathcal{O}}_{\alpha\beta}
\label{time_expectation_local_operators}
\end{align}
where $\hat{\mathcal{O}}_{\alpha\beta}=\langle E_{\alpha} | \hat{\mathcal{O}} | E_{\beta} \rangle$. For non-degenerate energy levels, it can be seen from the above equation that, the second term vanishes after a sufficiently long time and only the first term contributes, indicating the emergence of a steady state. In the case of a complex many-body system, ETH demands that the diagonal elements of the observable $\mathcal{O}$ acquire an equilibrium value $\bar{\mathcal{O}}$ which smoothly varies with the energy density $E$ and is independent of the basis vectors i.e. $\mathcal{O}_{\alpha\alpha} = \bar{\mathcal{O}}(E)$. It can be argued that, if the Hamiltonian resembles a random matrix with random eigenvectors, the average value of the observables acquires a constant value $\bar{\mathcal{O}}$, which corresponds to the microcanonical description. A more general ansatz was proposed by Srednicki \cite{Srednicki1999} which is stated as follows,
\begin{eqnarray}
\hat{\mathcal{O}}_{\alpha\beta} = \bar{\mathcal{O}}(\bar{E})\delta_{\alpha\beta} + e^{-\mathcal{S}(\bar{E})/2}f(\bar{E},\omega)\xi_{\alpha\beta}
\end{eqnarray} 
where $\bar{E}=(E_{\alpha}+E_{\beta})/2$, $\omega=E_{\alpha}-E_{\beta}$, $S(\bar{E})$ is the thermodynamic entropy at energy $\bar{E}$, and  $f(\bar{E},\omega)$ is a smooth function of its arguments. The random variables $\xi_{\alpha\beta}$ come from a Gaussian distribution with zero mean and unit variance. From the ensemble equivalence, the expectation value $\bar{\mathcal{O}}(\bar{E})$ corresponds to the microcanonical average at energy $\bar{E}$. Such correspondence of the expectation value of the observable for a generic eigenstate within a sufficiently small energy window around $\bar{E}$ to its microcanonical average at that energy is related to the issue of `typicality', which is extensively discussed in Ref.\cite{Goldstein2006,Short2006,Reimann2007,Reimann2008}.
Even in the presence of additional conserved quantities, thermalization in a many-body system can still be achieved for which the steady value of the observables correspond to that of a `generalized Gibbs ensemble' (GGE) \cite{Rigol-GGE,Rigol-GGE2,Caux-GGE,GGE_experiment_Langen2015}. Recently, the ETH has also been generalized for non-commuting conserved charges \cite{Non_Abelian_ETH}. 
Moreover, ongoing studies have unveiled structure beyond ETH, encompassing the statistical correlations between the matrix elements of the observables as well as the energy eigenstates \cite{Chalker_ETH,Kurchan_ETH_OTOC,Prosen_general_ETH,
Chalker_stat_properties}.
Although the validity of ETH has been tested for many interacting quantum systems \cite{Polkovnikov-review,Santos-review,Rigol2009,Kollath2010,Sorg2014,Ikeda2014,Gogolin2014,
Haque2014,Haque2015,Neuenhahn2012,Fratus2016,HamazakiETH2022,
Santos_Dicke2022,two_mode_Dicke}, a rigorous proof is still lacking. In fact, whether or not ETH is a necessary condition for thermalization is still a debatable issue \cite{Gemmer2015,Cramer2015,Mori2017,ETH_review_Deutsch2018}.

\section{Quantum classical correspondence and collective models}
\label{quantum_classical_correspondence_and_collective_models}
Generally, manifestation of the underlying phase space dynamics in its quantum counterpart can provide a better insight to understand the route to ergodicity \cite{Ullmo1993,Voros,Voros_lecture}. In the integrable regime, the semiclassical Einstein–Brillouin–Keller (EBK) quantization bridges the gap between classical dynamics and the corresponding eigenspectrum \cite{Wimberger_book}. However, it is a pertinent issue to understand the fate of the quantum system in the presence of perturbations that lead to the onset of classical chaos due to the breaking of invariant tori. 
Although the BGS conjecture provides an indirect link between the fully chaotic dynamics and eigenvalue statistics, such a connection is not fully understood for dynamical systems that undergo a transition to chaos (see for example Fig.~\ref{Fig1}).
The intermediate regime, especially when the regular dynamics coexists with the chaotic motion giving rise to a mixed phase space structure, can lead to intriguing implications on quantum ergodicity, which requires an involved analysis. Such correspondence can be explored in the models that offer an appropriate tunable semiclassical limit, such as the large spin collective systems, which we elaborate in this section.

\begin{figure*}
	\centering
	\includegraphics[width=\textwidth]{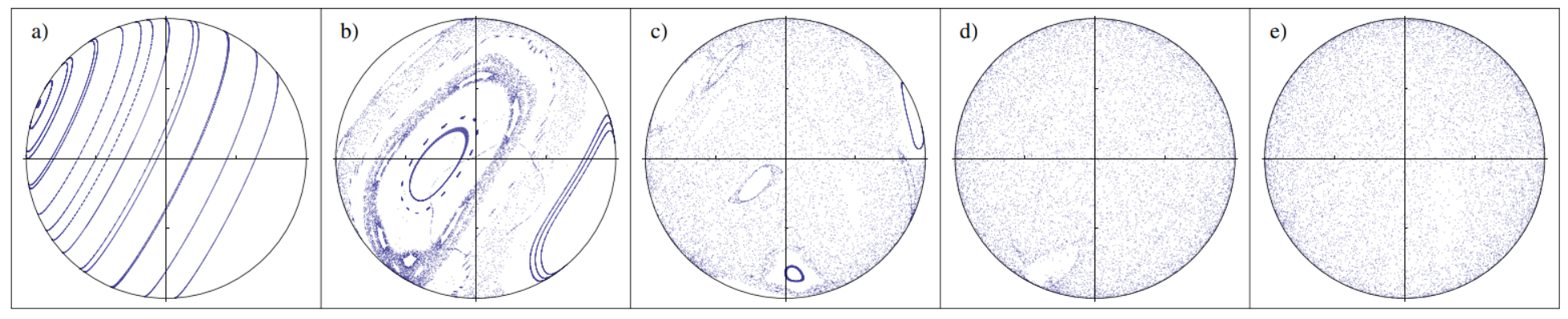}
	\caption{Poincar\'{e} sections generated by monitoring the pair $(s_{x},s_{y})$ at fixed values of oscillator variables $q$ and $p$, at energies $\Delta \epsilon = 30|\epsilon_{0}|$ above the ground state.  For each parameter value, $\gamma$, nine trajectories of different on-shell initial conditions are sampled. (a) $\gamma = 0.2$, (b) $\gamma = 0.7$, (c) $\gamma = 0.9$, (d) $\gamma = 1.01$, and (e) $\gamma = 1.5$. Here $\gamma = \lambda/\lambda_{c}$ with $\lambda_{c}=\sqrt{\omega\omega_{0}}/2$ and $\epsilon \equiv \mathcal{H}_{cl}/\omega_{0}$. In all the figures, only the projection of the southern hemisphere is shown, although the projection of the northern hemisphere looks qualitatively similar. Reprinted (figure) with permission from Ref.\cite{Altland_Haake_DM_PRL}, Copyright (2012) by the American Physical Society.}    
\label{Fig3}
\end{figure*}

A classic example of a collective model is the Dicke model (DM), which was originally introduced to describe the super-radiant phase in an atom-photon interacting system \cite{Dicke1954}. However, in recent years, extensive studies on this model have revealed several fascinating phenomena starting from quantum phase transition to the onset of chaos \cite{Emary_Brandes_DM, Emary_Brandes_DM-rev}. The DM describing a collection of $N$ two-level atoms in the presence of a radiation field is represented by the following Hamiltonian,
\begin{eqnarray}
\hat{\mathcal{H}} = \hbar\left(\omega_{0}\hat{S}_{z} + \omega \hat{a}^\dagger \hat{a} + \frac{2\lambda}{\sqrt{N}}(\hat{a}+\hat{a}^\dagger)\hat{S}_{x} \right)
\label{Dicke_model}
\end{eqnarray}
where $\hat{S}_{\alpha}$ ($\alpha=x,y,z$) denote the components of the large spin (describing the collection of $N$ two-level atoms) with magnitude $S=N/2$, and $\hat{a}(\hat{a}^\dagger)$ is the photon annihilation (creation) operator. The first two terms of the Hamiltonian describe the atoms and photons, where $\omega_{0}$ and $\omega$ denote the atomic excitation and photon frequency, respectively. The third term describes the interaction between them with coupling strength $\lambda$. This model undergoes a quantum phase transition at a critical coupling strength $\lambda_{c}=\sqrt{\omega\omega_{0}}/2$, above which a superradiant phase emerges, that is characterized by non-vanishing photon number \cite{Hepp_Lieb_QPT,Emary_Brandes_DM,Emary_Brandes_DM-rev}.  It also exhibits a range of fascinating phenomena that have a correspondence with their classical counterpart \cite{Altland_Haake_DM_PRL, Altland_Haake_DM_NJP, Brandes2013, Hirsch2014a, Hirsch2014b, atom_field_chaos2016, Hirsch2016}.

The semiclassical limit for the DM can be obtained by introducing the new scaled operators,
\begin{eqnarray}
\hat{s}_{\alpha} &=& \hat{S}_{\alpha}/S, \quad
\hat{q} = \frac{(\hat{a}+\hat{a}^\dagger)}{\sqrt{2S}}, \quad \hat{p}=\frac{\imath(\hat{a}^\dagger-\hat{a})}{\sqrt{2S}}
\end{eqnarray}
satisfying the commutation relations $[\hat{s}_{i},\hat{s}_{j}] = \imath\hbar\epsilon_{ijk}\hat{s}_{k}/S$ and $[\hat{q},\hat{p}]=\imath \hbar/S$, where $\hbar$ is reduced by a factor of $S$. Due to this, the collective systems such as the DM are appropriate to explore the correspondence with the underlying dynamics, since they provide a semiclassical description in the large $S$ limit. For $S \rightarrow \infty$, the commutators mentioned above will vanish and as a consequence, the scaled operators can be treated as classical variables $s_{\alpha}$, $q$, and $p$. In this limit, $s_{\alpha}=\vec{S}/S = (\sin{\theta}\cos{\phi},\sin{\theta}\sin{\phi},\cos{\theta})$ denotes the scaled components of a large spin vector $\vec{S}$, where $(\theta,\phi)$ represents its orientation on a unit Bloch sphere. On the other hand, $(q,p)$ corresponds to the dimensionless canonical coordinates of a harmonic oscillator. Alternatively, the spin variable can be represented by the canonical conjugate variables $\{\phi,z=\cos{\theta}\}$. In terms of the above variables, the classical limit of Hamiltonian in Eq.~\eqref{Dicke_model} is given by,
\begin{align}
\mathcal{H}_{cl}&=\left(\omega_{0}z + \frac{\omega}{2}(q^2+p^2) + 2q\lambda\sqrt{1-z^2}\cos{\phi}\right) 
\label{classical_Hamiltonian}
\end{align}
where the classical energy corresponds to the energy density $E/S$.
Another advantage of the large spin systems is that one can describe the phase space points semiclassically using the coherent states \cite{Radcliffe1971}, which is a minimum uncertainty packet around that point.
Both the spin and bosonic sectors can be represented semiclassically by the following coherent state description, 
\begin{subequations}
\begin{align}
\ket{\phi,z} &= \left( \frac{1+z}{2}\right)^{\!S}\exp\left\{\!\sqrt{\frac{1-z}{1+z}}e^{\imath \phi}\hat{S}_{-}\!\right\}\!\ket{S,S}\\
\ket{\alpha} &= \exp\left\{\alpha \hat{a}^\dagger-\alpha^{*}\hat{a}\right\}\ket{0}
\end{align}
\end{subequations}
where $\hat{a}|\alpha\rangle = \alpha|\alpha\rangle$ and  $\alpha = \sqrt{2S}(q+\imath p)/2$. The product state $\ket{\psi} = \ket{\phi,z} \otimes \ket{\alpha}$ is an appropriate representation of the phase space points and therefore serves as a suitable choice to probe the underlying phase space structure in the dynamics of the Dicke-like collective models, which is discussed extensively in Sec.~\ref{quantum_scarring_phenomena_in_collective_models}.

The phase space dynamics of the DM can readily be analyzed from Hamilton's equations obtained using Eq.~\eqref{classical_Hamiltonian}, which reveals various interesting features. When the interaction strength is sufficiently small, the DM exhibits regular motion. With increasing coupling strength $\lambda$, the quantum phase transition to the superradiant phase at the critical coupling strength $\lambda_{c}$ is manifested in the classical dynamics as a pitchfork bifurcation of the steady state corresponding to the ground state \cite{Emary_Brandes_DM}. Notably, above the transition point $\lambda_{c}$, the onset of chaos occurs and is reflected from the appearance of the chaotic sea, which surrounds the regular regions around the fixed points in the phase space. Eventually, the chaotic sea covers the full phase space for large $\lambda$. 
Such crossover from the regular to chaotic dynamical behavior has also been analyzed from the change in the spectral statistics from Poisson to Wigner-surmise \cite{Emary_Brandes_DM,Emary_Brandes_DM-rev}, as depicted for the Dicke model in Fig.~\ref{Fig2}. 
As evident from the case study of the DM, the degree of chaoticity can be tuned by the change in the system parameters, which in turn has a direct influence on ergodicity and thermalization.

\begin{figure}
	\centering
	\includegraphics[width=\columnwidth]{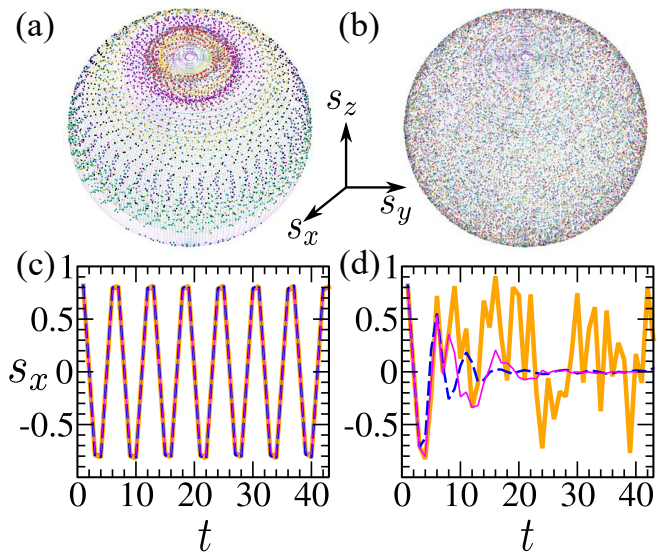}
	\caption{Classical quantum correspondence in the kicked Dicke model (KDM): Stroboscopic evolution on Bloch sphere for interaction strengths, (a) $\lambda = 0.05$ and (b) $\lambda = 0.6$. Comparison between classical, quantum, and truncated Wigner dynamics (thick solid orange, blue dashed, and thin pink lines, respectively) starting from the initial state $\{z=0.36,\phi=0.5,q=0.11,p=0.14\}$ for (a) $\lambda=0.05$  and (b) $\lambda=1.0$. Parameters chosen for quantum dynamics: $S=20$, $N_{\rm max}=30$ (Here $N_{\rm max}$ corresponds to  Fock space truncation). In all figures, $\omega_{0}=0.5$ and $T=\pi/3$. For more details, see also Ref.\cite{Ray2016}. (a) and (b) Reproduced from Ref.\cite{OTOC_KDM}. \textcopyright\, IOP Publishing Ltd. All rights reserved. (c) and (d) Reprinted (figure) with permission from Ref.\cite{Ray2016}, Copyright (2016) by the American Physical Society.} 
\label{Fig4}
\end{figure}

It is clear from the classical dynamics that, the onset of chaos triggers the phase space mixing. For time-independent DM, such mixing in the phase space is clearly reflected from the Poincar\'e section of the spin sector at a fixed energy. As shown in Fig.~\ref{Fig3}, the phase space mixing is enhanced with increasing interaction strength $\lambda$, which induces the thermalization process. From such a classical perspective, an alternate viewpoint of thermalization in the DM has been considered by Altland and Haake \cite{Altland_Haake_DM_PRL, Altland_Haake_DM_NJP}.
In this case, the semiclassical phase space density described by the Q-distribution evolves following a Fokker-Planck type diffusion equation, which facilitates thermalization to a microcanonical form in the chaotic regime. More recently, the validity of ETH has been tested for the DM \cite{Santos_Dicke2022,two_mode_Dicke}, which also confirms thermalization in the chaotic regime. Additionally, the accuracy of the microcanonical description increases for higher energy states, which suggests an energy-dependent ergodic behavior (see also the discussion in Sec.~\ref{energy_dependent_ergodicity}).  Another manifestation of thermalization is the emergence of a diagonal reduced density matrix. Interestingly, the correspondence between phase space mixing at a fixed energy density and the emergence of such diagonal ensemble has been elaborated for the coupled top model \cite{Sinha_CT2020,Sinha_CT2022}.

As an extension, a periodically driven Dicke model, called the `kicked Dicke model' (KDM) has been considered to investigate the transition to chaos and its manifestation on quantum dynamics and ergodic behavior \cite{Ray2016}. The time-dependent Hamiltonian describing this system is given by,
\begin{align}
   \hat{\mathcal{H}}(t) &= \omega_{0}\hat{S}_{z} + \Omega \hat{a}^\dagger \hat{a} + \frac{\lambda_{0}}{\sqrt{2S}}(\hat{a}+\hat{a}^\dagger)\hat{S}_{x}\!\! \sum^{\infty}_{n=-\infty}\!\!\delta(t-nT)
    \end{align}
where $\omega_{0}$ denotes the energy gap of the two-level atoms and $\Omega$ corresponds to the frequency of the cavity mode. The periodic drive is introduced in the coupling between the spin and the bosonic sectors, where $\lambda_{0}$ represents the kicking strength and $T$ is the time period between two consecutive kicks. Such periodic pulses were originally considered in the well-known `kicked top model', where the kicking is applied to a single large spin \cite{Haake1987,Haake2010}.
The KDM also has a well-defined semiclassical limit which enables one to obtain the stroboscopic dynamics. With increasing the kicking strength, this model exhibits a crossover to the chaos which can be evident from the phase portrait of both spin and oscillator degrees of freedom, see Fig.~\ref{Fig4} and also Ref.\cite{Ray2016}. 
As expected, for small kicking strength $\lambda_{0}$, there is a good agreement between the classical and quantum dynamics of the spin components, which is evident from Fig.~\ref{Fig4}(c). 
In the quantum counterpart, the signature of chaos is manifested in the quasienergy level spacing distribution, which changes from Poisson to Wigner surmise \cite{Ray2016}.

Unlike the DM, due to the continuous pumping of energy in KDM, such a Floquet system thermalizes to infinite temperature in the deep chaotic regime, which can be described by a diagonal reduced density matrix with equal weights, corresponding to the microcanonical picture. Consequently, the spin dynamics exhibits relaxation to a steady state, where the average of the spin components vanish (see Fig.~\ref{Fig4}(d)). As a reflection, during the time evolution, both the kinetic energy and the potential energy increase with equal proportion, giving rise to the equipartition of energy.
Note that, for numerical computations with finite bosonic number states, a deviation from the microcanonical picture is observed in the smaller subsystem, that is the spin sector, which exhibits the signature of the canonical thermalization in the presence of the bosonic bath.

Another variant of DM can be realized when a bosonic Josephson junction (BJJ) is coupled with an external (cavity) mode. Ultracold atoms loaded into a double well trap constitute a Bose Josephson junction, which exhibits coherent oscillation due to the tunneling of atoms between two wells \cite{BJJ_expt_oberthaler1, BJJ_expt_schmiedmayer, BJJ_expt_oberthaler2, BJJ_expt_levy, BJJ_expt_zibold, BJJ_walls, BJJ_shenoy1, BJJ_shenoy2, BJJ_leggett1, BJJ_tonel, BJJ_vardi3, BJJ_kroha1, BJJ_review}, similar to Josephson oscillation in superconductors. For a fixed number of atoms $N$, such BJJ can be described by the Lipkin-Meshkov-Glick (LMG) model \cite{LMG1,LMG2,LMG3} of large spin with magnitude $S = N/2$ \cite{BJJ_walls}. When the atoms in two wells of the BJJ are coupled with two external bosonic modes, this system can effectively be described by an extended Dicke-like model, given by \cite{Sinha_BJJ_dissipative,Sinha_BJJ_scars},
\begin{align}
&\hat{\mathcal{H}} = -J\hat{S}_{x} + \frac{U}{2S}\hat{S}^2_{z} + \frac{\lambda}{2\sqrt{2S}}\hat{S}_{z}(\hat{b}+\hat{b}^\dagger) + \omega_{0}\hat{b}^\dagger\hat{b}
\label{extended_Dicke_model}
\end{align}
where $J$ denotes the tunneling amplitude between two wells, $U$ is the onsite interaction strength, $\lambda$ represents the coupling between the spin and the bosonic sectors, and $\omega_{0}$ corresponds to the frequency of the external bosonic mode. Much like the DM, BJJ also undergoes a QPT, above which the onset of chaos occurs \cite{Sinha_BJJ_dissipative, Sinha_BJJ_scars}, that has also been analyzed from the eigenvalue as well as the eigenvector statistics in Ref.\cite{Wang_EDM}. 
This model is particularly interesting, as it exhibits a spectacular effect on the degree of ergodicity due to the presence of a dynamical state known as `$\pi$-mode' (see also Refs.\cite{BJJ_expt_zibold, BJJ_shenoy1, BJJ_shenoy2}), which can exist in the chaotic regime.

\begin{figure}
	\centering
	\includegraphics[width=1.025\columnwidth]{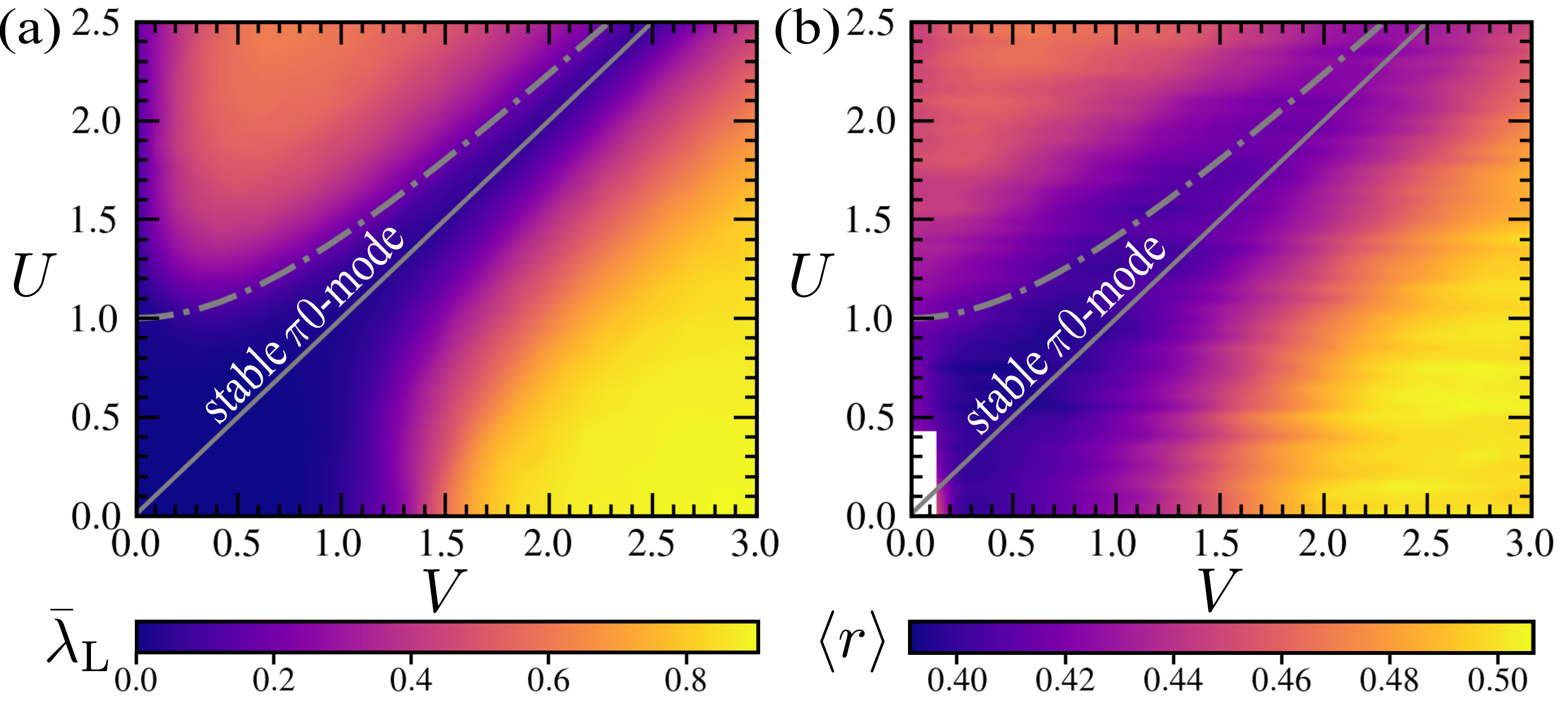}
	\caption{Classical quantum correspondence in the two-component BJJ in terms of ergodicity. The colormaps show the degree of ergodicity quantified from (a)  Lyapunov exponent $\bar{\lambda}_{\rm L}$ averaged over an ensemble of phase space points (classical chaos) and (b) average level spacing ratio $\langle r \rangle $ (quantum chaos) in the parameter space $(U,V)$. The steady state `$\pi0$-mode' is dynamically stable between the solid and dashed gray lines, whereas it is unstable otherwise. A small region near the origin is kept blank in (b) since it is difficult to get reliable statistics close to the non-interacting regime ($U=0$,$V=0$). Reprinted (figure) with permission from Ref.\cite{two_component_BJJ}, Copyright (2022) by the American Physical Society.}    
\label{Fig5}
\end{figure}

A similar situation can also be observed in BJJ of a binary mixture of bosonic atoms where different dynamical steady states exhibit dramatic influence on overall ergodicity \cite{two_component_BJJ}.
Within the Schwinger-boson representation, the BJJ formed by a mixture of bosons with an equal number of atoms $N$ of each species ($i=1,2$) can be described by two interacting large spins with magnitude $S=N/2$. The effective Hamiltonian is then given by,
\begin{align}
\hat{\mathcal{H}} = -J(\hat{S}_{1x}+\hat{S}_{2x})+\frac{U}{2S}\left(\hat{S}^2_{1z}+\hat{S}^2_{2z}\right)+\frac{V}{S}\hat{S}_{1z}\hat{S}_{2z}
\label{two_component_BJJ_ham}
\end{align}
where the spin components are given by $\hat{S}_{ix}=(\hat{a}^\dagger_{iL}\hat{a}_{iR}+{\rm h.c})/2$ and $\hat{S}_{iz}=(\hat{n}_{iL}-\hat{n}_{iR})/2$. In the above equation, $\hat{S}_{ix}$ denotes the tunneling of the $i$th species, while $\hat{S}_{iz}$ represents their population difference between the left (L) and right (R) well of the BJJ. The parameters $J$ and $U$ ($V$) correspond to the tunneling rate and the intra- (inter-)species interaction between the two components, respectively. For $U=0$, this model reduces to the coupled top \cite{Sinha_CT2020,Sinha_CT2022}.
In the case of two-component BJJ, the crossover to chaos depends on the interaction strengths $U$ and $V$. Particularly, in the absence of $V$, this model reduces to the integrable LMG model. Classically, the degree of chaoticity for different interaction strengths $U$ and $V$ can be quantified from the Lyapunov exponent $\bar{\lambda}_{\rm L}$ averaged over an ensemble of phase space points. On the other hand, in the quantum counterpart, the average level spacing ratio $\langle r \rangle$ has been computed in the $(U,V)$ plane. As illustrated in Fig.~\ref{Fig5}, the remarkable resemblance between these two quantities $\bar{\lambda}_{\rm L}$ and $\langle r \rangle$ over the parameter space unveils the classical-quantum correspondence in terms of ergodicity. From this example, it is evident that the classical dynamics has a significant influence on the overall ergodic behavior of the quantum counterpart, which also has an interesting consequence leading to the deviation from ergodicity that we discuss later in Sec.~\ref{deviation_from_ergodicity}.

The above-mentioned examples of collective quantum systems, particularly the Dicke-like model and its variants can also be realized in the cold atom \cite{Esslinger-rev2013, dicke_cavity_esslinger, Barrett2014, dicke_cavity_hemmerich, Barrett2017, Keeling-rev2019} and circuit quantum electrodynamics setups \cite{cQED2014}. Such systems exhibit a plethora of intriguing many-body phenomena. Additionally, these collective models have become an ideal platform to probe the impact of underlying phase space dynamics on quantum ergodicity, which we discuss in the next sections.

\section{Out-of-time-order-correlator (OTOC): an indicator of chaos}
\label{OTOC}
Dynamical chaos in a classical system is an intriguing phenomena which can give rise to the butterfly effect i.e. exponential sensitivity of the initial conditions. However the notion of butterfly effect in quantum systems is a bit ambiguous due to the absence of classical trajectories. Moreover, a quantitative measure akin to Lyapunov exponent is also important for revealing the underlying chaos in a many-body quantum system. Classically, the Poisson brackets of canonical conjugate variables $\{q,p\}$ can be used to compute the Lyapunov exponent $\lambda_{\rm L}$ related to the exponential sensitivity of the initial conditions,
\begin{equation}
    \{q(t),p(0)\}_{\rm P.B}=\frac{\partial q(t)}{\partial q(0)} \sim e^{\lambda_{\rm L} t}
\end{equation}
Analogously, to diagnose chaos in the corresponding quantum system, a quantity called Out-of-time-ordered correlator (OTOC) $C(t)$ has been introduced for suitably chosen operators $\hat{W}$ and $\hat{V}$, where the Poisson bracket is replaced by the commutator in the above relation,
\begin{equation}
    C(t) = {\rm Tr}\left(\hat{\rho}[\hat{W}(t),\hat{V}(0)]^{\dagger} [\hat{W}(t),\hat{V}(0)]\right)
\end{equation}
which can be computed for a suitable density matrix $\hat{\rho}$, both at zero and finite temperatures.
For the choice of the operators, $\hat{W}= \hat{q}$ and $\hat{V}=\hat{p}$, it is expected that OTOC grows exponentially with time as $C(t) \sim e^{2\lambda_{\rm L}t}$, which can be employed to detect chaos in a quantum system. 
The OTOC was originally introduced in the context of disordered superconductors \cite{Larkin}, and has regained interest after its application in the Sachdev-Ye-Kitaev (SYK) model \cite{SYK1,SYK2} which yields an upper bound to the Lyapunov exponent in connection with the thermalization of the black holes \cite{Shenker2014,Maldacena2016}. Recently, it has become widely popular as an alternate and more direct way to detect `many-body quantum chaos' \cite{Shenker2014, Maldacena2016, Swingle-QFT-OTOC, Galitski2017, Sachdev2017, Knap2017,Swingle_unscrambling2018, information_scrambling2020,information_scrambling_finite_temp,OTOC_anisotropicDicke, quantum_ising_chain2018,finite_speed_scrambling,OTOC_classicalquantumDicke, quantum_classical_crossover,OTOC_KDM,Swingle2016,Lev2017,Galitski-levstat,GarciaMata-OTOC,GarciaMata-OTOC-fluc,Richter_saturation,markovic_saturation,Arul-scrambling,Fradkin-OTOC,Ray-OTOC-MBL,Alet-OTOC,RMT_isospectral_twirling,Clifford_gates_quantum_chaos,p-spin_glass}. Moreover, in the context of scrambling phenomena, this method has its application in the quantum information theory as well \cite{Swingle2016, Lev2017, finite_speed_scrambling, Knap2017, quantum_ising_chain2018, information_scrambling2020, information_scrambling_finite_temp,RMT_isospectral_twirling}. In addition, the present day developments in experimental techniques to measure the quantum correlations enable a direct investigation of the OTOC in trapped ions and NMR systems \cite{OTOC_trapped_ion, OTOC_expt_NMR,  expt_verified_scrambling, expt_scrambling_quantum_circuits, OTOC_expt_hbc}. 

Another similar quantity $F(t)$ has also been utilized which is related to the quantum version of the butterfly effect \cite{Shenker2014,Swingle-QFT-OTOC, Swingle_unscrambling2018, information_scrambling2020},
\begin{eqnarray}
F(t)= {\rm Tr}\left(\hat{\rho}\hat{W}^\dagger(t)\hat{V}^\dagger(0)\hat{W}(t)\hat{V}(0)\right)
\end{eqnarray}
where $C(t)=2[1-F(t)]$ holds when $\hat{W}$ and $\hat{V}$ are unitary operators.

\begin{figure}
	\centering
	\includegraphics[width=1.02\columnwidth]{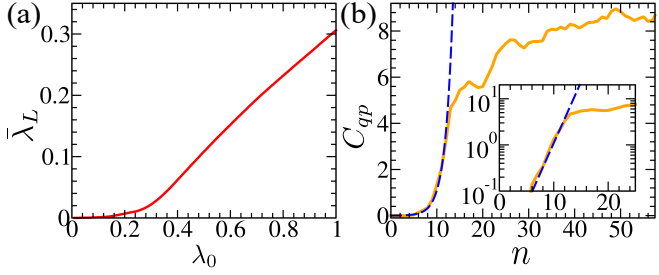}
	\caption{Classical and quantum chaos in KDM. (a) Variation of the maximum Lyapunov exponent (averaged uniformly over the phase space) with increasing kicking strength $\lambda_{0}$. (b) Stroboscopic dynamics of OTOC in the oscillator sector for $\lambda_{0}=1$ using an initial coherent state corresponding to a typical point in the classical phase space. The inset shows the same as a semi-log plot. The blue dashed line is generated from $\hbar^2_{\rm eff}e^{2\lambda_{\rm L}n}$, where $\lambda_{\rm L}$ corresponds to the classical Lyapunov exponent at $\lambda_{0}=1$ and $\hbar_{\rm eff}=1/S$ is the effective Planck's constant. Parameters chosen: $S=20$, $N_{\rm max} = 120$, $\Omega=0.5$, and $T=\pi/3$. Reproduced from Ref.\cite{OTOC_KDM}. \textcopyright\, IOP Publishing Ltd. All rights reserved.}   
\label{Fig6}
\end{figure}

Note that, the relation between the exponential growth rate of $C(t)$ and the classical Lyapunov exponent holds only for the systems with appropriate semiclassical limit. Such correspondence has been tested for single particle classically chaotic system by tuning the effective Planck's constant $\hbar_{\rm eff}$ \cite{Galitski2017}. Usually, for these systems, the OTOC grows exponentially between the time $t_{\rm d} \sim 1/\lambda_{\rm L}$ till the scrambling/Ehrenfest time $t_{\rm E}\sim 1/\lambda_{\rm L} \log (1/\hbar_{\rm eff})$ \cite{Maldacena2016,Galitski2017}, after which it attains a saturation due to the finite size of the system. 

In this context, the collective models are promising candidates to test the effectiveness of OTOC to diagnose chaos in an interacting quantum system. For example, the onset of chaos in KDM has also been investigated using the OTOC dynamics \cite{OTOC_KDM}. As demonstrated in Fig.~\ref{Fig6}, the OTOC constructed in the bosonic sector of the KDM grows exponentially  with time, and its growth rate increases with the kicking strength which controls the underlying chaoticity. 
It has been found that, the growth rate of $C(t)$ is in good agreement with the classical Lyapunov exponent in the chaotic regime. 
Moreover, the ergodic to non-ergodic transition by tuning the non-integrable perturbation in the anisotropic Dicke model has also been captured from the dynamics of $F(t)$ \cite{OTOC_anisotropicDicke}. Its striking similarity with the average level spacing ratio $\langle r \rangle$, which is a common indicator of chaos, highlights the potential of OTOC as a tool for detecting chaos in an interacting quantum system.

\begin{figure}
	\centering
	\includegraphics[width=1.02\columnwidth]{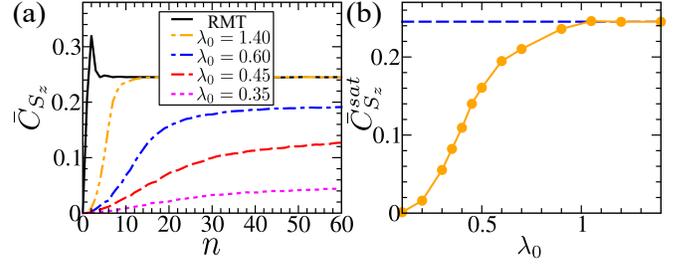}
	\caption{Saturation value of OTOC as a measure for the quantum signature of chaos in KDM. (a) Stroboscopic time evolution of the OTOC $\bar{C}_{S_{z}}(n) =-{\rm Tr}([\hat{s}_{z}(n),\hat{s}_{z}(0)]^2\hat{\rho}_{\rm mc})$, where $\hat{\rho}_{\rm mc} = \sum_{\nu}\ket{\psi_{\nu}}\bra{\psi_{\nu}}/\mathcal{N}$ is the microcanonical density matrix with $\ket{\psi_{\nu}}$ being the eigenstates of the Floquet operator of dimension $\mathcal{N}$ and $\hat{s}_{z}=\hat{S}_{z}/S$ is the scaled spin operator. (b) Variation of long time Saturation value $\bar{C}^{sat}_{S_{z}}\equiv \bar{C}_{S_{z}}(n\rightarrow\infty)$ with increasing kicking strength $\lambda_{0}$. The blue dashed line denotes the maximum value $\bar{C}^{\rm max}_{S_{z}} \sim \langle \hat{s}^2_{z} \rangle^2_{\rm mc}$ approached by $\bar{C}^{sat}_{S_{z}}$ in the limit of large $S$, where $\langle \hat{s}^2_{z} \rangle_{\rm mc} = {\rm Tr}(\hat{s}^2_{z}\hat{\rho}_{\rm mc})$. The parameters chosen are same as Fig.~\ref{Fig6}. Reproduced from Ref.\cite{OTOC_KDM}. \textcopyright\, IOP Publishing Ltd. All rights reserved.}   
\label{Fig7}
\end{figure}

Apart from the growth rate, the long time behavior of OTOC, especially its saturation value also turns out to be another marker for ergodic behavior in the chaotic regime \cite{GarciaMata-OTOC,GarciaMata-OTOC-fluc,OTOC_KDM,OTOC_anisotropicDicke, Galitski-levstat, Richter_saturation, Arul-scrambling,markovic_saturation,Ray-OTOC-MBL}. 
In the case of the KDM, the saturation value of the OTOC constructed from the spin operators captures the onset of chaos, as well as probes the degree of ergodicity for increasing kicking strength, which is summarized in Fig.~\ref{Fig7}. Moreover, in the disordered systems, the long-time dynamics and the saturation value of OTOC have also been used to understand the many-body localization to delocalization transition in the presence of periodic drives, see Ref.\cite{Ray-OTOC-MBL}.

In a recent study, a variant of OTOC, namely, the fidelity out-of-time-order-correlator (FOTOC) has been introduced, and using the paradigmatic Dicke model, it has been demonstrated that FOTOC can elucidate connections between information scrambling, entanglement, ergodicity and quantum chaos, see Ref.\cite{Rey-FOTOC}. In this method, the operators are chosen as, $\hat{W}=e^{i\delta \phi \hat{G}}$ and $\hat{V}=|\psi_0\rangle \langle \psi_0|$, where $\hat{G}$ is a Hermitian operator. The main advantage of FOTOC is that, it becomes proportional to the fluctuation of $\hat{G}$ corresponding to the initial state $\ket{\psi_{0}}$ when the parameter $\delta \phi$ is sufficiently small i.e. $\delta \phi \ll 1$. This measure has proved to be effective for detecting the underlying dynamical instability in the collective models \cite{Rey-FOTOC, Santos-FOTOC, Corney-FOTOC, Sinha-KCT}. Both the degree of ergodicity as well as its deviation has been probed by using FOTOC for coupled top \cite{Sinha_CT2022} and kicked coupled top \cite{Sinha-KCT}.

\begin{figure*}
	\centering
	\includegraphics[width=\textwidth]{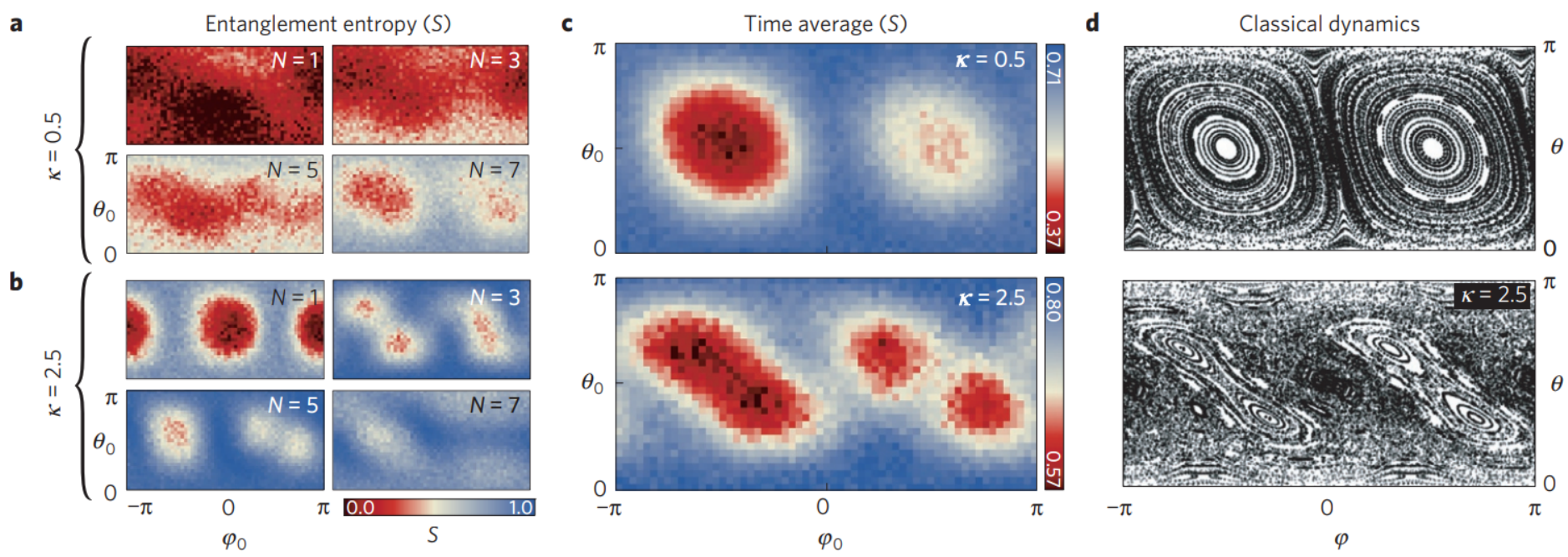}
	\caption{(a,b) Color plot of the entanglement entropy of a single qubit averaged over qubits and mapped over a $31 \times 61$ grid of the initial state, for various time steps and two values of interaction strength (a) $\kappa =0.5$  and (b) $\kappa =2.5$. (c) Entanglement entropy averaged over 20 steps for $\kappa =0.5$ and over 10 steps for $\kappa =2.5$; for both experiments, the maximum pulse sequence is $\approx 500$ ns. (d) Stroboscopic map of the classical dynamics computed numerically for comparison. The map is generated by randomly choosing $5000$ initial states, propagating each state forward using the classical equations of motion, and the orientation of the state is plotted after each step as a point. A clear connection between regions of chaotic behavior (classical) and high entanglement entropy (quantum) is observed. Reproduced from Ref.\cite{Neill2016} with permission from Springer Nature.}   
\label{Fig8}
\end{figure*}

A classical version of the OTOC, namely, the  `decorrelator' and its dynamics has also been investigated in order to study chaos as well as the butterfly effect in a classical many-body system \cite{Huse2018, Moessner2018}. In the case of a spin chain with three-dimensional classical spin vector $\Vec{S}_i$, the decorrelator is defined as,
\begin{equation}
    \mathcal{D}(i,t) = 1 - \langle \Vec{S}_i^a(t).\Vec{S}_i^b(t)\rangle,
\end{equation}
where $a$ and $b$ are two copies of spin configurations of the same initial state $(t=0)$, except a small initial perturbation in terms of spin rotation is applied to the spin at site $i_{0}$ corresponding to $b$. Here, $\langle . \rangle$ denotes the averaging over different initial spin configurations chosen randomly from an appropriate ensemble. To study the chaotic behavior of the system at finite temperatures, such ensemble corresponds to a thermal distribution \cite{Sthitadhi2022}, see Ref.\cite{Moessner2018} for details. The spatio-temporal growth of the decorrelator exhibiting a lightcone like spreading of the initial perturbation has also been observed for different spin models \cite{Moessner2018_1, Moessner2021, Sarang2021, Sthitadhi2022} as well as in other many-body systems \cite{Knolle2021,Subhro2021,Kulkarni2020,Sumilan2021}. 
Apart from the OTOC, in the recent years, the dynamical behavior of the spectral form factor (SFF) has also been used for the diagnosis of `many-body quantum chaos' \cite{SFF_analytical_PRX,exact_SFF_PRL,Chalker_SFF1,Chalker_SFF2}. 
Interestingly, the SFF has been obtained analytically for non-integrable periodically driven Ising spin chains, exhibiting its connection to RMT \cite{SFF_analytical_PRX,exact_SFF_PRL}.

Although, the OTOC and its variants discussed above have been used as a popular tool for the detection of `many-body quantum chaos', it is worth mentioning that, the growth rate of OTOC in certain cases may act as a false flag of chaos \cite{OTOC_false_positive2021}, and therefore should be treated carefully.
Interestingly, the exponential growth of OTOC and its variants have also been observed in the vicinity of the saddle points and unstable steady states of integrable models such as LMG, capturing the underlying instability \cite{Santos-FOTOC,scrambling_and_chaos2020,instability_rozenbaum2020,
scrambling_longrange_spinchains2018,scrambling_IHO,scrambling_coupled_harmonic_chains}.
Within this scenario, apart from its applicability in the chaotic systems, the OTOC can also act as a useful tool to investigate the non-ergodic behavior which stems from the dynamical instability, that we discuss in the later section (see Sec.~\ref{dynamical_signature}) for detection of scarring phenomena.

\section{Chaos and Entanglement}
\label{chaos_and_entanglement}
Quantum entanglement of a correlated many-body system remains a topic of great interest \cite{Fazio2008,Vidal_collectivesystems2007,Peschel2009,Horodecki2009}, one of the reasons being it contains much useful information about the ergodicity of an isolated quantum system. Typical measures that are of focus in this context are the bipartite entanglement entropy (EE), namely, von Neumann or $n$th R\'enyi EE, where the latter one is defined as \cite{Renyi-EE},
\begin{equation}
    S_n = \frac{1}{1-n}\log {\rm Tr}[\hat{\rho}_{\rm red}^n],
    \label{Renyi_EE}
\end{equation}
and the von Neumann EE is a variant of Eq.~\eqref{Renyi_EE} for $n\rightarrow 1$, which is given by,
\begin{equation}
    S_{\rm en} = -{\rm Tr}[\hat{\rho}_{\rm red}\log \hat{\rho}_{\rm red}],
    \label{von_EE}
\end{equation}
In Eq.~\eqref{Renyi_EE} and \eqref{von_EE}, $\hat{\rho}_{\rm red}$ is a reduced density matrix representing the subsystem $A$, which is a constituent of the full system $S=A\cup B$. It can be obtained by tracing out the degrees of freedom of subsystem $B$, namely, $\hat{\rho}_{\rm red}={\rm Tr}_{\rm B} \hat{\rho}$, where for an isolated system, $\hat{\rho}=|\psi\rangle \langle \psi|$ is a pure state constructed from the wavefunction $|\psi\rangle$ of the full system $S$. Apart from its application in quantum information theory, the scaling of EE with system size has relevance in characterizing the ergodic behavior of different many-body systems. Due to the extensive nature of the thermodynamic entropy, it is expected that, for ergodic systems, the EE follows a `volume law' scaling \cite{volumelaw_Rigol2022}, whereas, it has been observed that the systems which are reluctant to thermalize such as MBL phases, exhibit `area law' \cite{Abanin2019,Plenio2010}. Generally, the EE of a non-integrable many-body system possesses the thermodynamic characteristics analogous to that of a generic isolated system \cite{Deutsch_EE2010,Deutsch_EE2013}, which also hints towards a connection with thermalization.
In a seminal work, Page demonstrated that if a quantum system of Hilbert space dimension $\mathcal{N}=\mathcal{N}_{A}\mathcal{N}_{B}$ is in a random pure state, then the average entropy of a subsystem of dimension $\mathcal{N}_{A}$ (with $1 \ll \mathcal{N}_A \leq \mathcal{N}_{B}$) is given by \cite{Page-EE}, 
\begin{equation}
S_{\rm max} \simeq \log \mathcal{N}_A - \frac{\mathcal{N}_A}{2\mathcal{N}_{B}}
\label{page_value}    
\end{equation}
which sets an upper bound on EE and is referred to as the `Page value'. For non-integrable many-body systems, such a bound on EE of the eigenstates characterizing their ergodic nature and its connection with eigenstate thermalization is an involved topic of discussion \cite{Page-EE, Rigol-EE, Bianchi-EE, Srednicki-EE, Grover-EE}.

The dynamics of EE can also serve as a useful probe to characterize the nature of the system, namely, a logarithmic growth of EE, $S_{\rm en}(t) \sim \log t$, during a quench dynamics starting from a non-entangled state indicates a many-body localized regime, whereas, a linear growth of EE i.e. $S_{\rm en}(t) \sim t$ is an indication of ergodic dynamics of the system. For a detailed discussion on entanglement spreading during quench dynamics as well as its scaling behavior in different contexts, we refer to Ref.\cite{Plenio2010, Abanin2019}. Beyond the bipartite entanglement, a proposal for the measurement of multipartite entanglement came through computing the quantum Fisher information (QFI) \cite{Smerzi2012, Toth2012, Zoller2016, Toth2017}. In a recent study, the long time saturation value of the QFI has been shown to capture the regular-to-ergodic transition in the Dicke model \cite{Piazza2019}.

\begin{figure}
	\centering
	\includegraphics[width=\columnwidth]{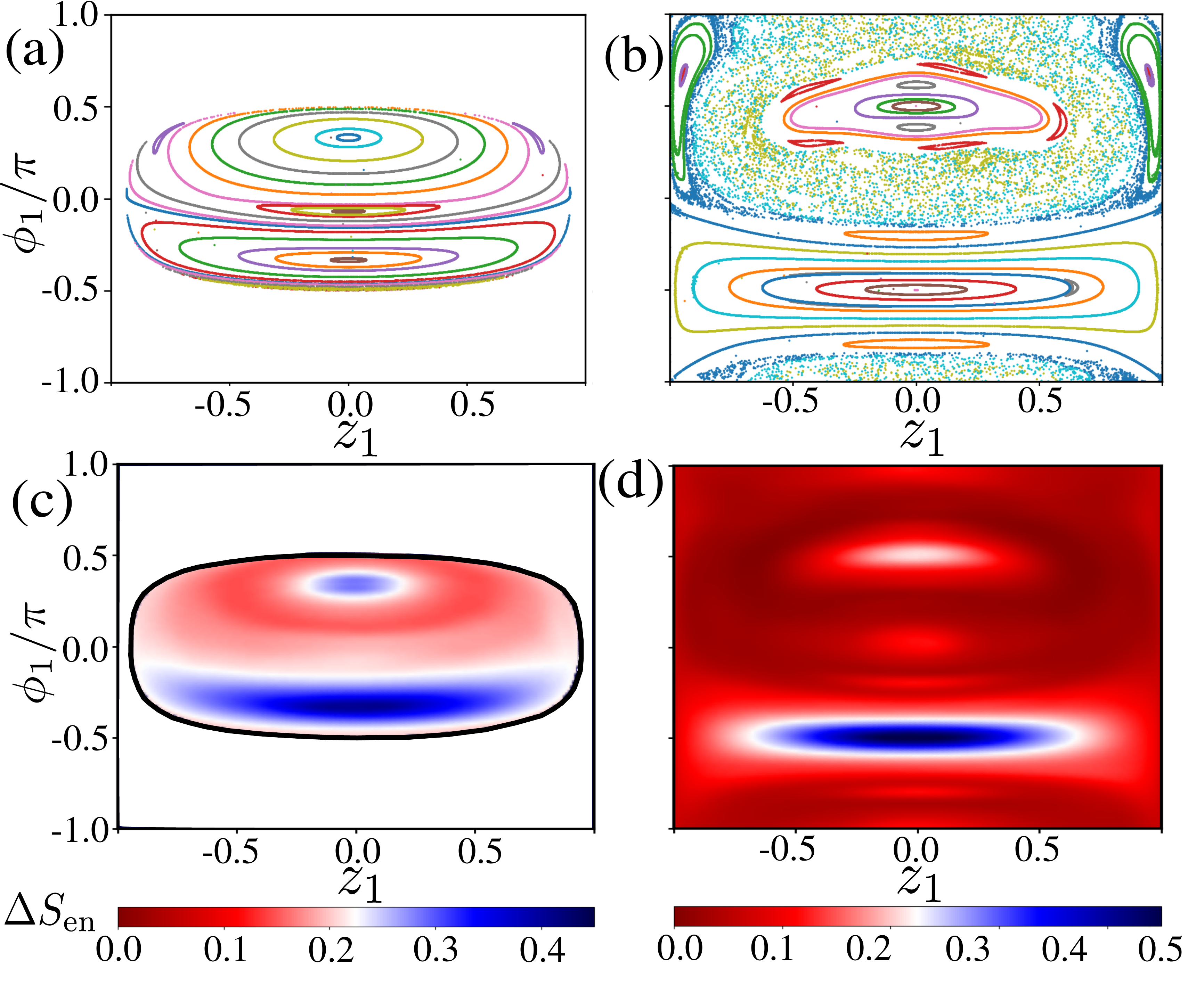}
	\caption{{\it Chaos, entanglement and energy-dependent degree of ergodicity:} Poincar\'e sections at $z_2 = 0$ plane for (a) $E = -1.0$ and (b) $E = 0.0$, for $U = 0.8$ and $V = 1.2$. Note that $\{z_{i},\phi_{i}\}$ with $i=1,2$ denotes the canonical conjugate variables for each component of the binary mixture in BJJ. (c), (d) Color-scaled plots of deviation $\Delta S_{\rm vN}$ of the time-averaged EE $\bar{S}_{\rm vN}$ from the Page value $S_{\rm max}$ in the ergodic limit for initial coherent states representing the same phase space points in panels (a) and (b), respectively. For quantum calculations, we set $S = 30$. Reprinted (figure) with permission from Ref.\cite{two_component_BJJ}, Copyright (2022) by the American Physical Society.}   
\label{Fig9}
\end{figure}

The EE can also be utilized as a tool to unveil the nature of the phase space dynamics. For example, in the KDM, when the kicking strength is small corresponding to the regular region of the phase space, the EE grows slowly, resulting in a good agreement between the classical and quantum dynamics \cite{Ray2016}. On the other hand, in the deep chaotic regime, the EE exhibits rapid growth and finally saturates to the maximum value which approaches the Page value. Physically, this phenomena is a reflection of the enhancement of EE due to the phase space mixing. 
In recent seminal experiments, the dynamical chaos has been diagnosed from entanglement dynamics for the kicked top model realized in the ultracold $^{133}$Cs atoms \cite{Chaudhury2009} and in superconducting qubits \cite{Neill2016}. Such striking similarity between the phase space chaos and time-averaged EE as observed in the latter experiment is shown in Fig.~\ref{Fig8}.
This connection between chaos and entanglement has also been investigated theoretically in detail in Ref.\cite{Lewenstein-QKT, Ghose-QKT, Lombardi-QKT, Pattanayak-QKT, Lerose-QKT}.
Interestingly, the mixed phase space behavior where the regular and the chaotic regions coexist, can also be probed from the entanglement dynamics. In the case of two-component BJJ, the EE of the time-evolved state starting from an initial coherent state is able to distinguish the regular and chaotic regions of the underlying mixed phase space \cite{two_component_BJJ}, as illustrated in Fig.~\ref{Fig9}. The saturation value of the entanglement dynamics can also reveal the connection between energy-dependent ergodic behavior and phase space dynamics restricted on a fixed energy surface, which is elucidated in the coupled top model \cite{Sinha_CT2022}. In a generic non-integrable system, the saturation value of EE in the steady states can be used to characterize the degree of ergodicity as well as the underlying chaos.

Apart from the entanglement growth followed by its saturation, the entanglement spectrum (ES) also contains interesting features related to the ergodic nature of a quantum state.  The ES represents the eigenvalues $\lambda_{\alpha}$ of the reduced density matrix,
\begin{equation}
    \hat{\rho}_{\rm red} = \sum_{\alpha} \lambda_{\alpha} |\alpha\rangle \langle \alpha|,
\end{equation}
which is obtained from the Schmidt decomposition of a quantum state, $|\psi\rangle = \sum_{\alpha} \sqrt{\lambda_{\alpha}} |\alpha\rangle_{\rm A} \otimes |\alpha\rangle_{\rm B}$. Here, $|\alpha\rangle_{\rm A/B}$ are the basis states constituting the Hilbert spaces of the subsystems.  The ES of an ergodic state is extended and the corresponding eigenvalues $\{\lambda_{\alpha}\}$ are distributed according to the `Marchenko-Pastur distribution' of the RMT \cite{Pastur1967}, which can be also observed in a strongly chaotic system \cite{Arul2002}. Moreover, the ES has also been used to characterize the non-ergodic behavior in the MBL phase \cite{powerlaw_MBL2016,powerlaw_MBL2017}. Thus, the features embedded in the ES can identify ergodicity as well as its breaking, the latter of which we will discuss in more detail in the upcoming sections.

\section{Deviation from ergodicity}  
\label{deviation_from_ergodicity}
Although it is expected that a complex system with many degrees of freedom can exhibit ergodicity in non-equilibrium dynamics, there are different routes of ergodicity breaking which are the subject of intense research in recent years. Lack of thermalization or athermal behavior has recently been observed in several models that are simulated in cold atom experiments, particularly one-dimensional systems such as quantum Newton's cradle \cite{Kinoshita2006}.
At the classical level generically, integrability and a large number of conserved quantities hinder chaotic dynamics and ergodicity. However, ergodicity breaking in a closed quantum system is more subtle, and apart from integrability the nature of eigenstates plays a crucial role in it. 
Mainly, strong breaking of ergodicity occurs in systems exhibiting `many-body localization' (MBL) which prevents ergodic evolution and thermalization \cite{Basko-MBL,MBL_quasiperiodic2013,Huse-rev-MBL, Abanin2019}. Similar to the Anderson localization \cite{Anderson-loc}, the MBL phase is usually induced by disorder and quasi-periodic potential, which lead to the localization of ground state as well as excited state wavefunctions resulting in an `athermal' non-ergodic behavior. Unlike the ergodic phase, Poissonian level statistics is a characteristic feature of MBL \cite{D_Huse_2007_MBL}. An important dynamical signature of MBL includes logarithmic growth of entanglement, which is in contrast to the linear behavior exhibited by an ergodic phase. However, in the presence of sufficiently strong interactions or time-periodic perturbations, such MBL phase can become ergodic \cite{MBL_quasiperiodic2013,Fradkin-OTOC,Ray-OTOC-MBL,drive_induced_AA2016,Rafael2017} which has also been demonstrated in recent experiments \cite{delocalization_MBL_Bordia2017}. Moreover, there can also exist non-ergodic extended (NEE) or critical states, that are neither completely localized nor extended, which exhibit the multifractal properties, as discussed in Sec.~\ref{eigenvector_statistics}. Generally, such states appear close to the mobility edge separating the localized and ergodic regions  
\cite{Luitz2015,ShankarDasSarma2015,drive_induced_AA2016,
Mondaini2019,Subroto2020,Sumilan2023}, 
at the critical point of Anderson transition \cite{Anderson_transitions_Mirlin2008,Altshuler2014}, in the Anderson model on disordered random regular graphs \cite{Altshuler2014,Altshuler2016}, across the MBL transition \cite{Santos2015,Santos_nonergodic2017,Nicolas2019}, and in disordered systems with long-range hopping \cite{powerlaw_hopping_deng2019}. In recent years, it has been observed that the ground states of certain many-body systems can also exhibit multifractal behavior \cite{multifractal_spinchains_Bogomolny2012,Lindinger2019}. The presence of such NEE states leads to anomalous behavior in the thermalization \cite{Luitz2016} as well as in the transport properties \cite{Luitz_transport2020,Subroto2020}, which is an active area of research.

\begin{figure}
	\centering
	\includegraphics[width=\columnwidth]{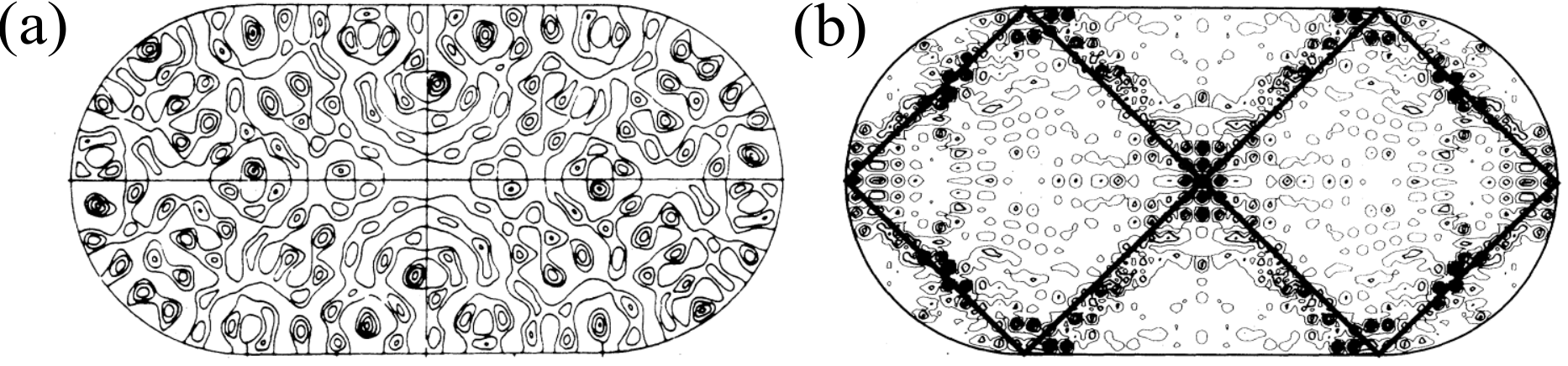}
	\caption{(a) Chaotic and (b) scarred eigenstates of the Bunimovich stadium. The solid lines in (b) denote the unstable periodic orbits corresponding to the scars. Only the negative contours are shown. Reprinted (figure) with permission from Ref.\cite{Heller1984}, Copyright (1984) by the American Physical Society.}      
\label{Fig10}
\end{figure}

Another source of weak ergodicity breaking has recently been observed in a remarkable experiment on ultracold Rydberg atoms, which surprisingly exhibited the revival phenomena for a particular choice of initial density wave (N\'{e}el-like) state, that is absent for arbitrary initial states in the ergodic regime \cite{Rydberg_expt_scar}. 
In an interacting many-body system, such revival dynamics occur due to the presence of certain `atypical' states in the ergodic regime, leading to weak ergodicity breaking, which have been coined as `many-body quantum scars' (MBQS) \cite{Abanin_Rydberg_scar2018, Papic_Rydberg_scar2018, Abanin_scars_review2021, Moessner_scars_review2022, Moudgalya_review2022}. 
Here, the weak breaking of ergodicity refers to strong dependence of the relaxation dynamics on the initial states in the ergodic regime \cite{Abanin_scars_review2021}. The memory of the initial states is retained during the dynamics due to the presence of a vanishingly small fraction of atypical which exhibit athermal behavior, since their statistical properties differ from the predictions of RMT.
In the context of deviation from ergodicity, the MBQS are analogous to {\it quantum scars} of non-interacting classically chaotic systems, that have been originally identified as reminiscent of the unstable classical orbits \cite{Heller1984,Kaplan1998,KaplanPRE1999,Kaplan1999,Heller2019}. For example, in the Bunimovich stadium, the scars of underlying classical trajectories are clearly visible from the nodal structure of higher energy wavefunctions (see Fig.~\ref{Fig10}) \cite{Heller1984}. 
Such scarring phenomena can also be viewed as localization of certain eigenstates \cite{Heller_PRA,Doron0,Doron1}. The scars can manifest as concentrations of probability around periodic orbits in the phase space. However, in the case of MBQS, such underlying classical structure is not apriori evident, which deserves further attention.

The above-mentioned experiment on Rydberg atoms can be well described by a `PXP model' \cite{Abanin_Rydberg_scar2018,Papic_Rydberg_scar2018} where the local projection operators `P' induce local constraints on the Hilbert space, that physically mimic the phenomena called `Rydberg blockade' \cite{ryd_blockade2000,ryd_blockade2001} such that two Rydberg atoms cannot sit next to each other. Here, `X' denotes the $x$-component of the local spin-1/2 operator representing the Rydberg excitations. It turns out that even in the ergodic regime, such kinematic constraints can give rise to athermal states that remain isolated from the thermal ones and are protected by an emergent symmetry \cite{Choi_Emmergent_Symmetry2019}. This kind of symmetry is usually associated with a `spectrum generating algebra' (SGA) \cite{SGA_book}, which is applicable to a subspace of the entire Hilbert space. Such algebra and its variants are associated with the mechanism behind the formation of a tower of low entangled athermal scarred states, which has been extensively studied for Affleck-Kennedy-Lieb-Tasaki (AKLT) chains \cite{AKLT_scar_Moudgalya1,AKLT_scar_Moudgalya2,AKLT_scar_Motrunich2020,AKLT_scar_MPS2020} and the spin-1 XY model \cite{spin1_XY_scars2019,spin1_XY_scarsMPS2020}.  
In the case of the spin-1 XY model, these towers of scarred states can be identified as bimagnon excitations \cite{spin1_XY_scars2019}, which invoke another picture of scarring where the scarred states can be considered as long-lived quasi-particles within the ergodic states \cite{Moessner_scars_review2022}. 
Interestingly, the scarred eigenstates of the PXP model can be interpreted as $\pi$-magnon excitations in a spin chain \cite{magnonscars2020}. 
Aside from SGA, the scarred states can also be constructed by systematically embedding non-thermal states within the thermal ones \cite{scar_Mori2017,AKLT_Shiraishi2019}. 
The tower of scarred states has also been found in different many-body systems such as in the Hubbard models \cite{eta_pairing_Moudgalya2020,eta_pairing_Motrunich2020}, bilayer systems \cite{bilayer_scars2022}, disordered spin chains \cite{Onsager_scars2020}, frustrated Kagome lattice \cite{kagome_scars2020}, transverse-field Ising ladder \cite{transverse_ising_ladder2020}, Heisenberg clusters \cite{Heisenberg_clusters2023}, amongst other examples \cite{ scars_deformedalgebra2020,rainbow_scars2022,multiple_magnons2022}.

More recently, another mechanism for ergodicity breaking known as `Hilbert space fragmentation' \cite{Pollmann2020,Motrunich_PRX2022,non_ergodicity_tilted_FHM2021,Moudgalya_review2022}, which is also related to `Hilbert space shattering' \cite{Nandkishore_fragmentation2020} and `Krylov fracture' \cite{Krylov_fracture_book}, has gained a lot of interest where the Hilbert space gets shattered into an exponential number of dynamically disconnected sectors. The intricate relationship between MBQS and weak fracturing of the Hilbert space has been explored in Bose-Hubbard models with correlated hopping \cite{correlated_hopping_BHM2020}, tilted 1D Fermi-Hubbard model \cite{tilted_1d_FHM2023}, optical lattices \cite{scars_optical_lattice2020}, spin chains \cite{Fradkin_chains2022, Kitaev_chain2022}, and random circuit models \cite{Fracton_scars2020}, for which the local constraints play a crucial role. In this context, it is interesting to note that suppression of chaos has also been observed in classical systems due to dynamical dynamical constraints \cite{Sthitadhi2022}.  

Another characteristic feature of the scarring phenomena is the the accumulation of the phase space density near the unstable classical orbits which has been observed in non-interacting systems.
In some sense, this bears resemblance with the quasi-particle picture of MBQS, where the scarred states can be viewed as non-decaying excitations surrounded by the ergodic states. To gain a better insight into such an intuitive description of scarring phenomena and its connection with the underlying dynamics, the collective quantum systems turn out to be appropriate candidates to explore, since they have a well-defined semiclassical limit. 
In the next subsection, we discuss the scarring phenomena and elucidate its connection with the underlying dynamics by considering a few examples of collective models that were introduced in Sec.~\ref{quantum_classical_correspondence_and_collective_models}.

\subsection{Quantum scarring phenomena in collective models}
\label{quantum_scarring_phenomena_in_collective_models}
In classical Hamiltonian systems, the transition from regular to chaotic dynamics occurs gradually by increasing the system parameters, as seen from the evolution of the phase space of the kicked rotor with increasing the kicking strength $\kappa$ (see Fig.~\ref{Fig1}). The dynamical behavior in the intermediate regime (of the kicking strength) exhibits regular islands that are surrounded by the chaotic sea, giving rise to a mixed phase space structure. Such coexistence of regular trajectories embedded in the chaotic sea bears an analogy to the appearance of the symmetry-protected athermal states or long-lived quasi-particles within the higher energy ergodic states, leading to the formation of MBQS. The mixed phase space structure is fascinating for its local chaotic behavior and its manifestation in the quantum counterpart requires a deeper investigation.

\begin{figure}
	\centering
	\includegraphics[width=1.025\columnwidth]{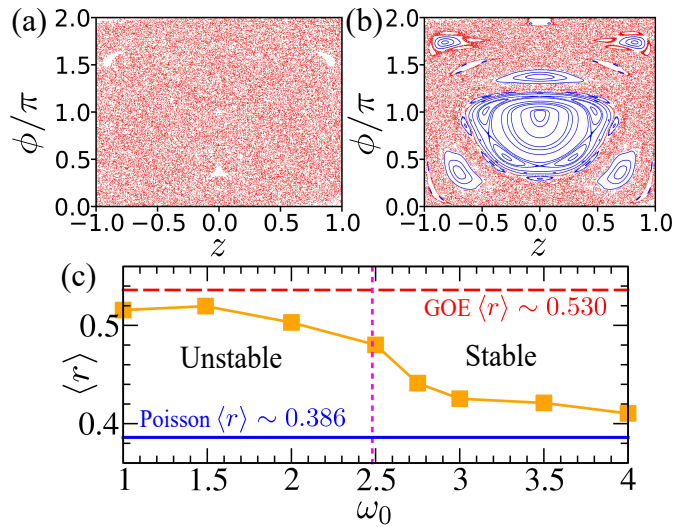}
	\caption{Impact of the stability of $\pi$-mode on the overall ergodic behavior of Bose-Josephson junction (BJJ) coupled to bosonic mode. Poincar\'{e} sections ($q=0$) at energy density $E \sim J$, exhibiting (a) chaotic behavior for $\omega_{0}=2.0$ and (b) mixed phase behavior at $\omega_{0}=3.0$. The red (blue) regions denote the chaotic (regular) parts of the phase space. (c) Variation of the average level spacing ratio $\langle r \rangle$ (see Eq.~\eqref{avg_level_spacing_ratio}) with frequency $\omega_{0}$ of the bosonic mode, indicating a crossover from chaotic to regular behavior across the instability point $\omega^{u}_{0}=2.48$. In all figures: $J=1.0$, $U=0.5$, $\gamma=1.2\gamma_{c}$, where $\gamma=\lambda^2/\omega_{0}$ and $\gamma_{c}=4(U+J)$. The pink dotted line denotes the instability point $\omega^{u}_{0}=2.48$, below which the $\pi$-mode becomes unstable. The solid blue (red dashed) lines denote the RMT values for the regular (chaotic) limits. For more details, see Ref.\cite{Sinha_BJJ_scars}. (c) Reprinted (figure) with permission from Ref.\cite{Sinha_BJJ_scars}, Copyright (2020) by the American Physical Society.}   
\label{Fig11}
\end{figure}

In an interacting collective system, one can obtain several higher energy steady states in addition to the ground state, from the semiclassical analysis. Such higher energy steady states in the chaotic regime can have very interesting implications on ergodicity, and as an example, we consider the case of BJJ coupled to a bosonic mode described by an extended Dicke model (EDM) given in Eq.~\eqref{extended_Dicke_model}. In particular, we focus on the effect of one of the steady states of EDM, known as `$\pi$-mode', which also arises as $\pi$ oscillations in the context of BJJ, when the average relative phase between the condensates is $\pi$ \cite{BJJ_expt_zibold, BJJ_shenoy1, BJJ_shenoy2}. This mode lies within the higher energy states, and its stability can be controlled by tuning the frequency of the bosonic mode, which in turn influences the dynamical behavior. When this mode is stable, the poincar\'{e} sections reveal mixed phase space behavior, where a regular island is formed within the chaotic region surrounding this stable fixed point, as shown in Fig.~\ref{Fig11}(b). On the contrary, when it becomes unstable by tuning the frequency, the stable island gradually disappears and the whole region is covered by the chaotic trajectories (see Fig.~\ref{Fig11}(a)). Interestingly, in the quantum counterpart, the stability of such mode reduces the average level spacing ratio $\langle r \rangle$ from the GOE limit, indicating a decrease in the overall degree of ergodicity (see Fig.~\ref{Fig11}(c)). A similar effect has also been observed in the two-component BJJ \cite{two_component_BJJ}, where the stability of the `$\pi0$-mode' has a dramatic influence on the overall ergodicity, as shown in Fig.~\ref{Fig5}. In addition, as a consequence of the stability of such modes, a dip in entanglement entropy (EE) can also be observed for the states corresponding to the energy density of the steady state. Such deviation of EE from its ergodic limit (Page value) in a generic many-body system is a characteristic feature of MBQS \cite{Abanin_Rydberg_scar2018,Papic_Rydberg_scar2018}. Remarkably, even when the dynamical steady states become unstable, their reminiscence can still be identified from certain eigenstates close to the corresponding energy densities. The statistical properties of such states like the Shannon entropy $S_{\rm Sh}$ and Inverse participation ratio $I_{q}$ can also exhibit significant deviation from the random matrix prediction. It is worth noting that ETH relies on Berry's conjecture, which states that the components of the high energy eigenstates of a quantum chaotic system in a generic basis follow Gaussian distribution \cite{Berry-conjecture}. In accordance with this conjecture, the scaled eigenstate components $\mathcal{N}|\psi_{i}|^2$ (with $\mathcal{N}$ being the Hilbert space dimension) of such higher energy states (that are typically ergodic) follow the `Porter-Thomas' (PT) distribution corresponding to a random eigenvector of the GOE matrix \cite{Haake2010}. In this context, the scarred eigenstates violate Berry's conjecture due to their athermal nature \cite{Sinha_BJJ_scars,Sinha_CT2020,Sinha_CT2022}. Such deviation from Berry's conjecture has also been reported in many-body systems leading to anomalous thermalization \cite{Luitz2016}.

\begin{figure}
	\centering
	\includegraphics[width=1.025\columnwidth]{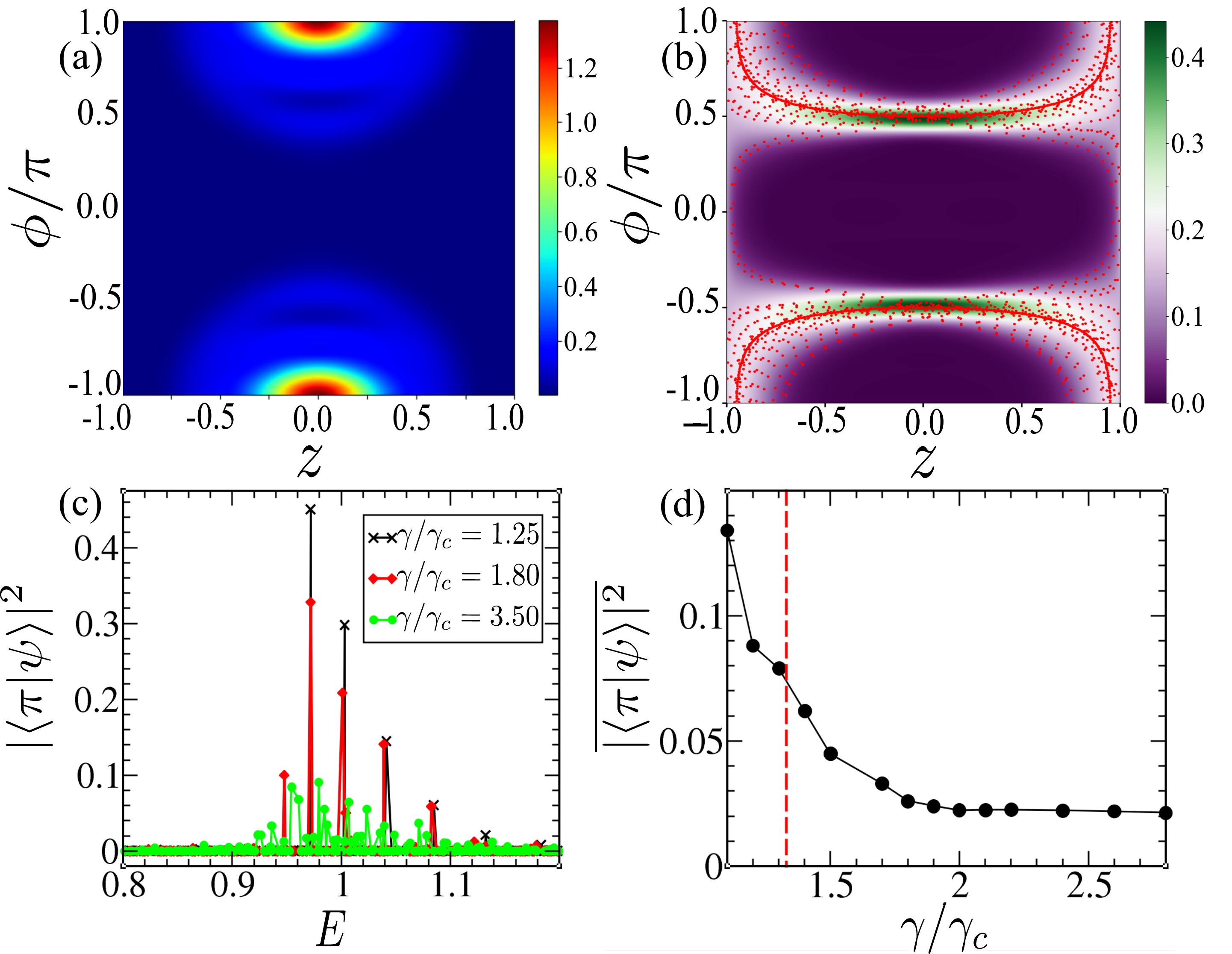}
	\caption{Quantum scar of the unstable dynamics in the extended Dicke model (see Eq.~\ref{extended_Dicke_model}) and the two-component BJJ (see Eq.~\ref{two_component_BJJ_ham}). Husimi distribution of the eigenstates containing the scar of the (a) unstable $\pi$-mode in EDM and (b) unstable periodic orbits in the two-component BJJ. The red dotted line in (b) denotes the classical trajectory starting from a phase space point in the vicinity of the unstable periodic orbit. (c) Overlap $|\langle \pi| \psi \rangle|^2$ of eigenstates with the coherent state $|\pi \rangle$ corresponding to the $\pi$-mode. (d) Variation of the average overlap $\overline{|\langle \pi |\psi \rangle|^2}$ with increasing coupling $\gamma/\gamma_{c}$, indicating the decreasing degree of scarring. The vertical dashed line denotes the instability of the `$\pi$-mode' at $\gamma/\gamma_{c}=1.33$. (a), (c), and (d) Reprinted (figure) with permission from Ref.\cite{Sinha_BJJ_scars}, Copyright (2020) by the American Physical Society. (b) Reprinted (figure) with permission from Ref.\cite{two_component_BJJ}, Copyright (2022) by the American Physical Society.}      
\label{Fig12}
\end{figure}

Moreover, from the Husimi distribution of such deviated states, a direct signature of scar can be identified through the accumulation of the phase space densities around the unstable fixed points or unstable orbits. For instance, when the $\pi$-mode becomes unstable in EDM, its reminiscence can clearly be identified from the localization of phase space density around the classical fixed point, as depicted in Fig.\ref{Fig12}(a,b). 
The eigenstates bearing the scars of the unstable steady state have a significant overlap with their semiclassical representation in terms of the coherent states, as evident for the $\pi$-mode in Fig.~\ref{Fig12}(c). On the contrary, the overlap of a random state is $~O(1/\mathcal{N})$ (with $\mathcal{N}$ being the Hilbert space dimension), which is smaller in comparison to the overlap of the scarred states \cite{Abanin_PRX2020}. When approaching the deep chaotic regime, the instability exponent of the corresponding unstable state increases, which is reflected in a decrease in the average overlap of the scarred states (see for example Fig.~\ref{Fig12}(d)). This can serve as a measure of the degree of scarring, which is analogous to the `quantum leakage' discussed in the context of the PXP model \cite{Lukin_Periodic_Orbits2019,Abanin_PRX2020}.
In the case of the two-component BJJ, as mentioned in Sec.~\ref{quantum_classical_correspondence_and_collective_models}, the overall ergodicity is suppressed by the stable $\pi0$-mode (see Fig.~\ref{Fig5}), which also exhibits similar scarring phenomena in the unstable regime \cite{two_component_BJJ}. In addition, the scar of the unstable periodic orbits can also be observed in these systems (see Fig.~\ref{Fig12}(b)) as well as the coupled top model \cite{Sinha_CT2020,Sinha_CT2022}. With increasing the interaction strength, the instability of the underlying steady state also increases, due to which, the degree of scarring decreases as the number of scarred states reduces (see for example Fig.~\ref{Fig12}(d) in the context of the EDM).
Interestingly, the restoration of quantum ergodicity in the Dicke model has been reported even when most of the eigenstates contain scar \cite{ubiquitous_scarring}. Strong evidence of MBQS has also been theoretically observed in a 1D homogeneous Bose-Hubbard model in the absence of any dynamical constraint, where scarring occurs in the vicinity of the unstable classical mean-field configurations \cite{Richter2023}.

Even in the presence of a periodic drive, the scarring phenomena have also been identified for interacting many-body systems, particularly in the driven PXP model \cite{PXP_drive_Krishnendu2020,PXP_drive_Saito2021,PXP_drive_Kawakami2020,PXP_drive_Papic2022} and other Floquet systems \cite{Doron2,DTC_scar2022,Fracton_scars2020}. 
In this context, deviation from ergodicity has also been analyzed for a strongly driven Floquet system due to the dynamical freezing and scar points \cite{Moessner_drive2021}. 
The signature of the unstable dynamics in the form of scar has also been studied in periodically driven large spin systems such as kicked top and kicked coupled top model \cite{Sinha-KCT}, kicked Dicke model \cite{OTOC_KDM} etc. Notably, the reminiscent of the unstable period 2-cycles (which are shortest periodic orbits in stroboscopic dynamics) has been investigated both in the kicked top and kicked coupled top \cite{Sinha-KCT}. This observation in the driven system motivates a search for a quantum analog of unstable periodic dynamical structures such as n-cycles.

\begin{figure}[b]
	\centering
	\includegraphics[width=\columnwidth]{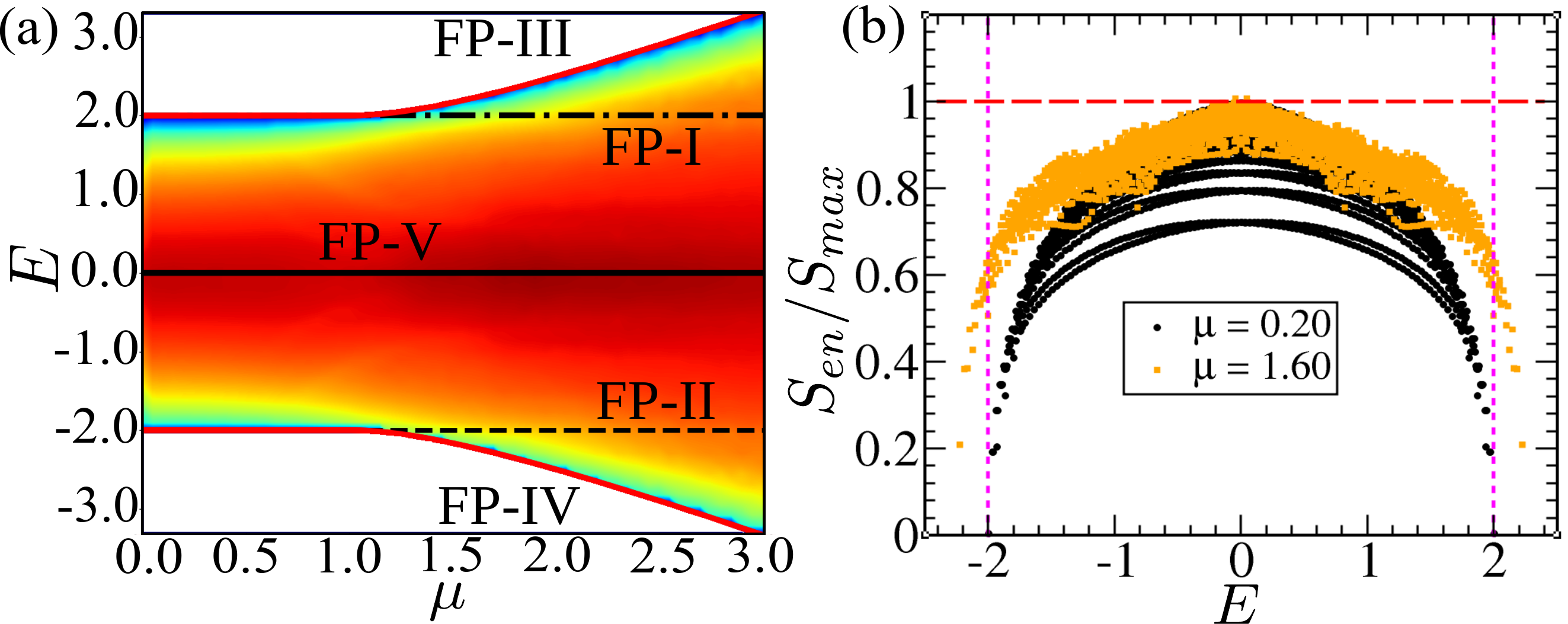}
	\caption{Energy-dependent ergodic behavior of the coupled top: Variation of the relative entanglement entropy (EE) $S_{en}/S_{max}$ (color plot) of eigenstates with energy densities $E$ for increasing coupling strength $\mu$. Variation of relative EE of eigenstates with energy density $E$ for different couplings $\mu$. The horizontal red dashed line represents the maximum EE at unity. Here, FP denotes the fixed points obtained from the classical analysis (for more details, see Ref.\cite{Sinha_CT2022}). Reprinted (figure) with permission from Ref.\cite{Sinha_CT2022}, Copyright (2022) by the American Physical Society.}      
\label{Fig13}
\end{figure}

\subsection{Energy-dependent ergodicity}
\label{energy_dependent_ergodicity}
As discussed in the previous section, the presence of an integrability breaking term in classical systems can lead to a mixed phase space structure with the coexistence of regular and chaotic dynamics, indicating local chaoticity. Similar behavior can also be observed when the dynamics is constrained over a fixed energy surface and the available phase space is restricted, giving rise to energy-dependent ergodic behavior, which also has implications in the quantum counterpart, particularly in the non-equilibrium dynamics. Even in a generic quantum chaotic system, all energy eigenstates are not uniformly ergodic i.e. the higher energy states or states that lie in the middle of the energy band with finite bandwidth are more ergodic compared to the ones that are close to the edge of the spectrum. The statistical properties of the eigenstates related to Berry's conjecture are a key ingredient of ETH, which is typically valid for higher energy states. In contrast, the states at the intermediate energy scales away from the band center exhibit deviation from the ergodic limit. The degree of ergodicity of different eigenstates can be quantified from the entanglement entropy (EE), which also turns out to be a good measure to probe local chaos \cite{Lewenstein-QKT, Ghose-QKT, Lombardi-QKT, Pattanayak-QKT, Lerose-QKT}. For example, the collective spin models with bipartite structure also exhibit such variation of the degree of ergodicity (in terms of the relative EE $S_{\rm en}/S_{\rm max}$) with energy density across the energy band, as shown in Fig.~\ref{Fig13} for the coupled top model \cite{Sinha_CT2020}. In the classically chaotic regime of this model, the EE at the band center approaches the Page value ($S_{\rm max}$) corresponding to a random state, however, it deviates significantly, away from the band center.

In addition, statistical properties of the eigenstates related to their non-ergodic behavior can be captured from the finite size scaling behavior of Shannon entropy $S_{\rm Sh}$ and generalized inverse participation ratio $I_{q}$, as discussed in Sec.~\ref{eigenvector_statistics}. In the coupled top model, the scaling of the $I_{q}$ with system size reveals the multifractal properties of the eigenstates away from the band center. Interestingly, in this model, the multifractal exponent $D_{1}$ and $S_{\rm en}/S_{\rm max}$ exhibit a linear relationship for eigenstates with increasing energy density. Such behavior of the multifractal dimension with EE has also been explored for non-ergodic `sparse random pure states' in Ref.\cite{Khaymovich2020}. 

Furthermore, strongly interacting quantum many-body systems can also display intriguing energy-dependent ergodic behavior \cite{Luitz2015,ShankarDasSarma2015,Mondaini2019,Subroto2020,
Buchleitner2021,Sumilan2023}. For example, in the case of interacting bosons in a quasi-periodic lattice, one can distinguish between localized, non-ergodic, and ergodic phases as the energy density increases, which can be captured from the EE \cite{ShankarDasSarma2015}. Such non-ergodic to ergodic transitions have also been observed in other systems \cite{powerlaw_hopping_deng2019}.

\subsection{Dynamical signature of scarring and energy-dependent ergodicity}
\label{dynamical_signature}
Since quantum scarring phenomena is closely related to the underlying dynamics, we first discuss how such local phase space structure can be probed in the collective quantum systems. As the coherent states are suitable semiclassical representations of the classical phase space points, the time evolution of these states can reveal the fingerprints of the local chaotic behavior. The time evolution of an ensemble of coherent states sampled uniformly over a region of phase space can unveil the mixed phase space structure from the distribution of EE as well as the Shannon entropy of the final states. For example, in the two-component BJJ, the time-averaged deviation of the EE from its ergodic limit reveals the underlying mixed phase space structure, where the maximum deviation of EE corresponds to the regular islands, as shown in Fig.~\ref{Fig9}. This method can also be applied to the driven systems such as kicked top and kicked coupled top, which reveals the underlying classical phase space structure, particularly the fixed points, and orbits from the quantum dynamics \cite{Sinha-KCT}.

\begin{figure}
	\centering
	\includegraphics[width=\columnwidth]{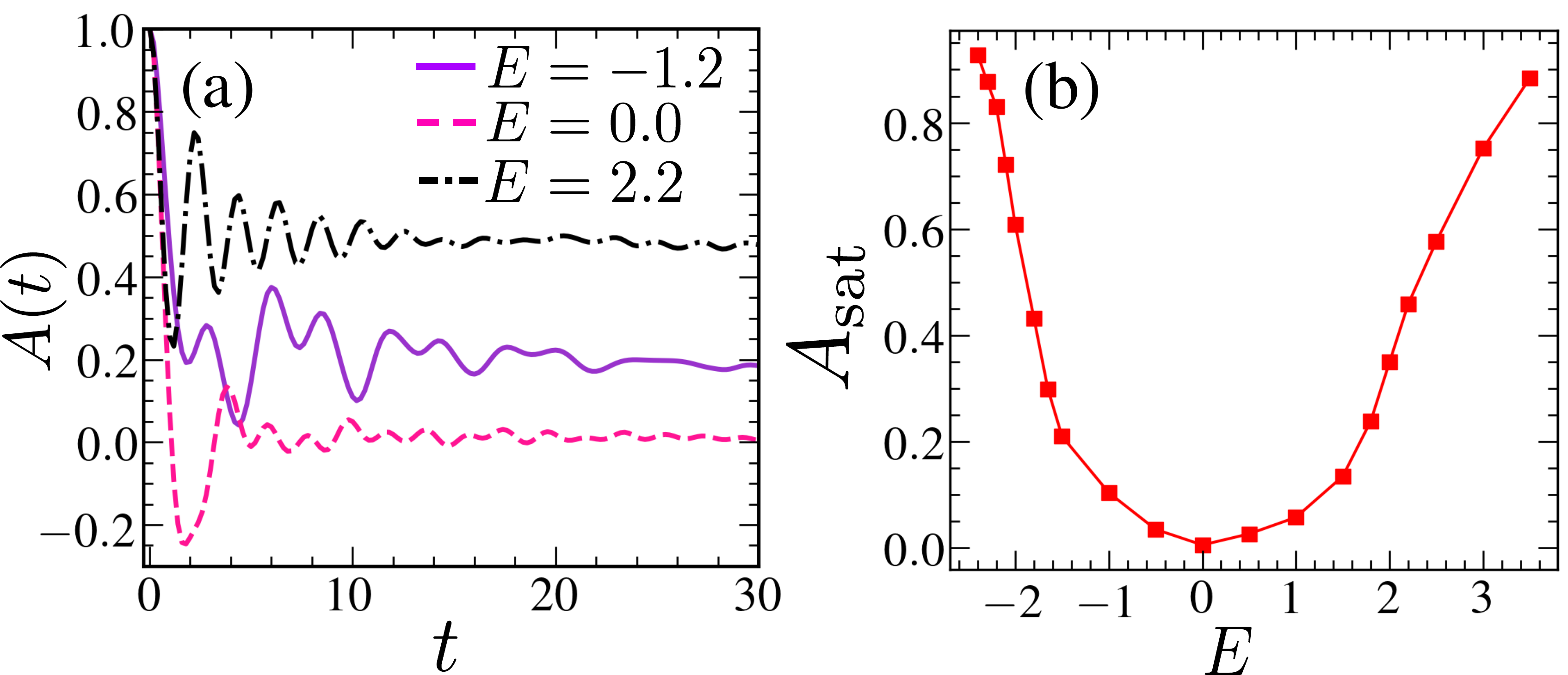}
	\caption{ Dynamical probing of the energy-dependent degree of ergodicity in the two-component BJJ: (a) Dynamics of auto-correlation function $A(t)$ evaluated at different energy densities $E$ and for $U = 0.8$, $V = 2.8$. (b) Variation
of saturation value of the auto-correlation $A_{\rm sat}$ across the energy spectrum. Reprinted (figure) with permission from Ref.\cite{two_component_BJJ}, Copyright (2022) by the American Physical Society.} 
\label{Fig14}
\end{figure}

Even when a stable island disappears in the chaotic sea of the phase space, its reminiscence in the form of `scar' can also be investigated by using the similar method of evolving the coherent state corresponding to such unstable fixed points. The quantum dynamics of such a special initial state exhibits athermal behavior, and the Husimi distribution of the time-evolved state displays the signature of `scar' in the form of semiclassical phase space density localized around the unstable fixed point. The signature of the athermal dynamics can be observed from the revivals of survival probability $F(t) = |\langle \psi(t)|\psi(0) \rangle|^2$, describing the Fidelity of the time-evolved state, which in contrast decays rapidly for an arbitrary initial state in the ergodic regime \cite{Sinha_BJJ_scars}. The dynamics of fidelity can also be used to detect the `scars' in the driven system, such as the scar of unstable 2-cycles (which is the smallest periodic orbit) in the kicked top as well as kicked coupled top \cite{Sinha-KCT}.
In the context of BJJ, the athermal behavior of the scarred states can be captured from the fluctuations of the relative phase between the condensates in the two wells, as the phase coherence is retained in the quantum dynamics \cite{Sinha_BJJ_scars}. On the contrary, for the thermal states, the phase fluctuations are rapidly enhanced, leading to the destruction of the phase coherence. 
Such dynamics of the phase fluctuation can be relevant, as it has been measured experimentally \cite{BJJ_expt_oberthaler2} (see also \cite{BJJ_review}).
The auto-correlation of relevant operators can also serve as a measure to detect the deviation from ergodicity. In recent years, the non-ergodic behavior of different quantum many-body systems has been investigated from the saturation value of auto-correlation functions \cite{Pollmann2020}. As discussed before, certain eigenstates of the collective models, such as two-component BJJ contain the `scar' of unstable periodic orbits, which can also be identified from the auto-correlation of the spin operators, 
\begin{eqnarray}
A(t)=\sum_{a=x,y,z} \langle\hat{S}_{1a}(t)\hat{S}_{1a}(0)\rangle
\label{auto_correlation_spins}
\end{eqnarray} 
The Fourier transform of such correlator evaluated for such scarred states exhibits a peak at a frequency, that corresponds to the period of the unstable classical orbit \cite{two_component_BJJ}. Moreover, the energy-dependent degree of ergodicity of two-component BJJ can also be probed from the saturation of such auto-correlation function (see Fig.~\ref{Fig14}) as well as from the phase fluctuations, exhibiting a complementary behavior across the energy spectrum.

In recent years, the out-of-time-ordered correlators (OTOC), introduced in Sec.~\ref{OTOC} have been used to detect the many-body quantum chaos, the growth rate of which plays an analogous role to the Lyapunov exponent in classical dynamics. However, OTOC has also been used to investigate the deviation from ergodicity, particularly in systems exhibiting MBL \cite{Ray-OTOC-MBL}. In the case of the collective systems, both OTOC and its variants like Fidelity-OTOC (FOTOC) have also been used to probe the degree of ergodicity as well as scarring phenomena. The OTOC evaluated for coherent state representing a particular phase space point turns out to be a suitable indicator of underlying local chaotic behavior. For the scarred states of the unstable fixed points, the OTOC exhibits slow growth and a small saturation value compared to an initial coherent state in the ergodic regime \cite{OTOC_KDM}. Furthermore, in the coupled top model, the time evolution of the OTOC for the eigenstates containing the scar of the unstable periodic orbits exhibits periodic oscillations with oscillation frequency close to that of the unstable orbit \cite{Sinha_CT2020}. From such oscillatory behavior in the FOTOC dynamics, the scarring phenomena have also been investigated in the periodically driven kicked coupled top \cite{Sinha-KCT}. 
Additionally, the saturation value of the FOTOC averaged over an ensemble of initial coherent states at a particular energy density is able to capture the energy-dependent ergodic behavior of the coupled top \cite{Sinha_CT2022}. Classically, the fixed energy density imposes a constraint on the accessible phase space of the system, which can be also reflected in the quantum dynamics.  The dynamical manifestation of such energy-dependent behavior can also be observed from EE as well as from the emergence of a diagonal reduced density matrix of the time-evolved state, which has been elucidated in Fig. 10 of Ref.\cite{Sinha_CT2022}.  

In summary, using the above-mentioned tools, the non-equilibrium dynamics of quantum many-body systems not only probe the degree of ergodicity but can also identify its deviation due to the presence of the athermal states, particularly the scarred states.

\section{classicality and entanglement spectrum}
\label{classicality_and_entanglement_spectrum}

\begin{figure}
	\centering
	\includegraphics[width=1.01\columnwidth]{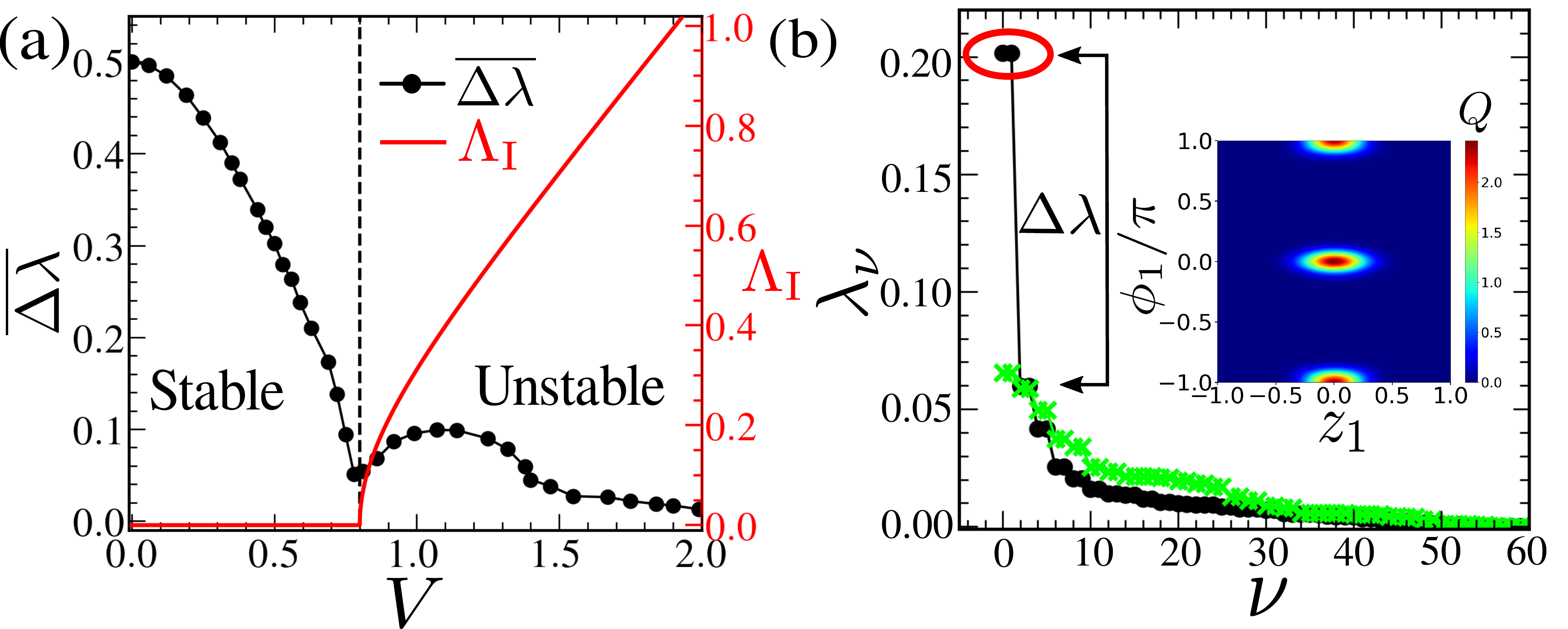}
	\caption{Quantum scar of `$\pi0$-mode' in the two-component BJJ: (a) Variation of time-averaged eigenvalue gap $\overline{\Delta \lambda}$ in the entanglement spectrum (left axis) and instability exponent $\Lambda_{\rm I}$ scaled by $J$ (right axis) are shown for the `$\pi0$-mode' with increasing interaction strength $V$. (b) Entanglement spectrum (ES) of eigenstate containing the scar of the `$\pi0$-mode' mode exhibiting a gap $\Delta \lambda$ (black circles), and of an arbitrary ergodic eigenstate (green crosses). In the inset, Husimi distribution $Q(z_{1},\phi_{1})$ of the reduced density matrix $\hat{\rho}^{\rm tr}_{S}$ for one of the spin sectors corresponding to largest eigenvalues marked by the red circle for the scarred eigenstate. Reprinted (figure) with permission from Ref.\cite{two_component_BJJ}, Copyright (2022) by the American Physical Society.}  
\label{Fig15}
\end{figure}

Underlying classicality in quantum dynamics and quantum scarring phenomena can be unfolded from the entangled spectrum (of the time-evolved state). This can be elucidated for the `$\pi0$-mode' of two-component BJJ, which is described by two interacting large spins. Such large spin systems with bipartite nature are suitable for analyzing entanglement and its connection with classical dynamics.   
In the regular regime of dynamics, a classical-quantum correspondence is reflected in the entanglement spectrum of the reduced density matrix of time-evolved state, which contains one or few large eigenvalues. On the other hand, in the chaotic regime, the entanglement spectrum corresponding to a typical ergodic state exhibits a continuous variation of eigenvalues which follows the `Marchenko–Pastur' distribution in accordance with RMT. In an intermediate regime, the entanglement spectrum can be divided into two parts, a few large eigenvalues that are separated by a gap from the tail of eigenvalue distribution. The semiclassical phase space density obtained from the effective density matrix constructed from these few large eigenvalues retains the underlying classical structure, whereas the long tail of eigenvalue distribution is related to the ergodic nature of the quantum state.  The gap in the entanglement spectrum can be used as a marker for the degree of underlying classicality in quantum dynamics. 
In the case of `$\pi0$-mode', the dynamical instability is manifested by the appearance of a dip in the variation of the gap in the entanglement spectrum \cite{two_component_BJJ}. Interestingly, this gap persists even after the instability of `$\pi0$-mode', retaining the memory of this unstable FP leading to the formation of scar (see Fig.~\ref{Fig15}). However, this gap decreases as the instability exponent increases, and finally, it vanishes in the deep chaotic regime where the entanglement spectrum behaves as that of a typical ergodic state. This phenomena is also related to the degree of scarring. 
Typical scarred states of these semiclassical systems exhibit such a gap in the entanglement spectrum, which has also been observed for scarred states of other interacting spin systems without any semiclassical picture \cite{Heisenberg_clusters2023}. 
Moreover, this feature is promising for unveiling classicality in a generic interacting quantum system by using product or matrix product states of low dimensionality \cite{Lukin_Periodic_Orbits2019, Abanin_PRX2020}.

\section{Discussion and outlook}
\label{discussion_outlook}

This review has aimed to offer a comprehensive overview of quantum chaos, ergodicity, and its deviation, as well as the latest developments in these areas of research. 
Quantum ergodicity, a concept that initially drew attention in the early days of quantum mechanics, has experienced a resurgence in interest, particularly due to significant advancements in the cold atom experiments.
Such a platform offers to study a variety of fascinating non-equilibrium phenomena \cite{krishnendu_colloqium, Langen_noneq_review}, and thus, it has opened new avenues for research in this direction. As an example, the eigenstate thermalization hypothesis (ETH) was proposed to shed light on the thermalization of isolated quantum systems, however, its deviation has already been observed in the experiments. Therefore, the regime of validity of such a hypothesis in a generic interacting quantum system is not always clear. This opens up the possibilities of exploring such phenomena from different angles. 

To understand the ergodicity of an isolated system, an alternate point of view, namely the phase space mixing triggered by the onset of chaos, has been explored in the collective quantum system called the Dicke model. A similar mechanism in a driven system, for example in the kicked Dicke model, can lead to the emergence of a diagonal density matrix, which describes a microcanonical thermalization in the deep chaotic regime. Such a semiclassical picture is advantageous for understanding the manifestation of local phase space behavior on the overall ergodicity of the quantum system, especially when the regular islands coexist with the chaotic sea. Interestingly, even in the chaotic regime, the presence of a stable steady state can have a dramatic impact on the overall ergodicity of the system, as has been observed in specific examples, namely in the extended Dicke model and in the two-component Bose Josephson Junction, where the ergodic dynamics is strongly suppressed due to the presence of a stable `$\pi$-mode'. In this context, the coherent states play an important role in unveiling the local phase space structure, retaining the underlying classicality during time evolution. The mixed phase space behavior may shed light on the deviation from the microcanonical formulation of thermalization, as well as from Berry's conjecture on the random nature of the eigenstates, however that requires further investigation.

Furthermore, in the mixed phase space regime, the gradual disappearance of the regular islands gives rise to the instability of the steady states, which in turn results in the formation of the quantum scars. Such scarring phenomena in the eigenstates of the collective quantum systems is clearly evident from the accumulation of semiclassical phase space density surrounding the unstable fixed points or periodic orbits. 
Even in the driven models such as kicked top and kicked coupled top, a similar scarring behavior due to the smallest unstable periodic orbits called 2-cycles can be probed from the stroboscopic dynamics.
On the contrary, such apriori connection with the unstable dynamics in the formation of many-body quantum scars (MBQS) in a generic interacting quantum system is missing. In this case, certain states are protected by an emergent symmetry, that gives rise to athermal dynamics. The dynamical constraints like the projection operators in the `PXP' model play a crucial role in the formation of MBQS, which are a subject of intense research \cite{Abanin_Rydberg_scar2018,Papic_Rydberg_scar2018, Abanin_scars_review2021, Moessner_scars_review2022}. 
Remarkably, the use of time dependent matrix product states (MPS) for the PXP model has revealed a mixed phase space structure corresponding to the scarring phenomena \cite{Abanin_PRX2020}. 
This approach seems promising for understanding an equivalent classical picture behind the formation of MBQS in a generic many-body system. Moreover, studies in the two-component BJJ reveal that the underlying classicality, as well as the formation of scars, can be captured from the appearance of a few large eigenvalues in the entanglement spectrum, which has also been observed in other spin systems \cite{Heisenberg_clusters2023}. Therefore, it is tempting to use low dimensional MPS to explore the underlying classicality of a generic many-body system in terms of low entanglement, in analogy with the coherent states of a collective large spin system.

Such semiclassical correspondence is always very encouraging due to which, the Out-of-time-ordered correlators (OTOC) have become a popular tool to detect chaos in a many-body system, also known as many-body quantum chaos. The dynamics of OTOC starting from an initial coherent state can probe the local chaotic structure, as well as the scarring phenomena. Although the growth rate of OTOC corresponds to the Lyapunov exponent in the semiclassical limit, its saturation value also serves as an important measure of the degree of ergodicity in an interacting quantum system.  
Similar to the Lyapunov exponent, which is a universally accepted measure of classical chaos, the pursuit of an analogous quantity to quantify the extent of ergodicity in interacting quantum systems still persists. While various metrics, such as entanglement entropy, OTOC, and spectral statistics, are frequently employed to probe ergodic behavior, a unifying measure remains a subject of interest.

The examples of the collective models considered in the present review not only serve as a suitable framework to investigate the link between ergodicity and the underlying dynamics, but can also be realized experimentally. For instance, the Bose-Josephson junction has already been implemented in the cold atom setups \cite{BJJ_expt_oberthaler1, BJJ_expt_schmiedmayer, BJJ_expt_oberthaler2, BJJ_expt_levy, BJJ_expt_zibold, BJJ_review}, for which the phase fluctuation between the condensates has been measured \cite{BJJ_expt_oberthaler2}, that can be used for the detection of the scarring phenomena. Large spin systems such as the kicked top model have also been emulated using cold atoms \cite{Chaudhury2009} and superconducting qubits \cite{Neill2016}. Likewise, the Dicke model as well as its variants can also be realized in the cavity and circuit quantum electrodynamics (QED). In this context, the recent experiments with Bose-Einstein condensates (BEC) coupled to the cavity modes are promising, which exhibit interesting non-equilibrium phenomena \cite{Esslinger-rev2013, dicke_cavity_esslinger, Barrett2014, dicke_cavity_hemmerich, Barrett2017, Keeling-rev2019}. Moreover, the circuit QED setups have also become a convenient platform for realizing such collective models \cite{cQED2014}. In these systems, the dissipation is inevitable due to various loss processes, because of which dissipative dynamics can also be probed, that can lead to the interesting possibility of exploring chaos and non-linear phenomena in a dissipative environment \cite{Lesanovsky2014,Nori2018,Parkins2020,Cohen2022}. Finding the fingerprint of chaos in such open quantum systems has also become an emerging area of research. The dissipative dynamics are described by non-unitary evolution, for which the Random Matrix Theory has been extended to non-Hermitian complex matrices \cite{ProsenPRX2020,Hamazaki2020,GarciaGarcia2022}. There are attempts to capture the departure from integrability in these open quantum systems from the spectral properties of the non-Hermitian matrices \cite{ProsenPRL2019}.
Another avenue to explore thermalization, entanglement growth, and non-ergodic behavior lies in the advancement of the random unitary circuits \cite{Bertini2021, CiracPRL2021, ClaeysPRL2021}. This is also important in the context of quantum information theory, where the time evolution can be performed by the application of the random unitary gates \cite{Fisher2022, ZollerPRX2022}.

Ergodicity is a widely recognized concept in statistical mechanics, yet it is not fully understood and has expanded the horizon for its exploration, particularly in open quantum systems. Furthermore, it has become equally important to identify the different pathways of its violation and their manifestations in real physical systems.

\section*{Acknowledgement}
We would like to pay our gratitude and respect to our colleague, late Prof. Amit Dutta, who had always been a great inspiration during many of the works. We thank our collaborators, particularly Debabrata Mondal and Johann Kroha. SR thanks Amichay Vardi and Doron Cohen for insightful discussions during collaborations on similar projects, which has helped in understanding the subject area to a great extent. SR acknowledges financial support by the Deutsche Forschungsgemeinschaft (DFG) under Germany’s Excellence Strategy—Cluster of Excellence Matter and Light for Quantum Computing, ML4Q (No. 390534769) and through the DFG Collaborative Research Center CRC 185 OSCAR (No. 277625399). SR also acknowledges funding by the Alexander von Humboldt (AvH) Foundation.

\end{document}